\documentclass[12pt]{article}

\usepackage{amsmath,amsfonts,latexsym,fullpage,graphicx,vmargin,natbib}
\usepackage{amssymb, mathrsfs}
\usepackage{pstricks}
\usepackage{subcaption}
\usepackage{amsmath}
\usepackage{multirow}
\usepackage{comment}
\usepackage[inline]{enumitem}
\usepackage{hyperref}
\usepackage[normalem]{ulem}
\usepackage{setspace}
\usepackage{textcomp}
\usepackage{booktabs}
\usepackage{placeins}
\usepackage[group-separator={,},group-minimum-digits={3}]{siunitx}

\newcommand{\iid}{\stackrel{\mbox{\scriptsize iid}}{\sim}}
\newcommand{\ind}{\stackrel{\mbox{\scriptsize ind}}{\sim}}
\newcommand{\bm}[1]{\mbox{\boldmath{$#1$}}}

\newcommand{\R}{\mathbb{R}}

\renewcommand{\mid}{\,|\,}

\usepackage{bm}

\newcommand{\e}{\mathrm{e}}

\newcommand{\MBtext}[1]{{\color{blue}{#1}}} 

\newcommand{\virgolette}[1]{``#1''}

\title{JAGS, NIMBLE, Stan: a detailed comparison among Bayesian MCMC software}

\author{Mario Beraha$^{1,2}$, Daniele Falco$^1$  and Alessandra Guglielmi$^{1}$\footnote{Corresponding author: Alessandra Guglielmi, alessandra.guglielmi@polimi.it}
	}
	
\date{
$^1$Department of Mathematics, Politecnico di Milano \\%
    $^2$Department of Computer Science, Universit\`a di Bologna\\[2ex]%
\today
}	
	
\begin{document}
	\maketitle
	
	\begin{abstract}
	The aim of this work is the comparison of  the performance of  the three popular software platforms  JAGS, NIMBLE and Stan. These  probabilistic programming languages are able to automatically generate samples from the posterior distribution of interest using MCMC algorithms, starting from the specification of a Bayesian model, i.e. the likelihood and the prior. The final goal is to present a detailed analysis of their strengths and weaknesses  to statisticians or applied scientists. In this way, we wish to contribute to make them fully aware of the pros and cons of this software.
We carry out a systematic comparison of the three  platforms on a wide class of models, prior distributions, and data generating mechanisms. 
Our extensive simulation studies evaluate the quality of the MCMC chains produced, the efficiency of the software and the goodness of fit of the output. 
We also  consider the efficiency of the parallelization made by the three platforms. 
	\end{abstract}


\section{Introduction}
\label{sec:intro}
	
Bayesian statistics offers a natural framework to quantify the uncertainty associated to statistical inference. Moreover,  it is straightforward to include prior knowledge into the model, for instance through information from previous experiments.
In the Bayesian framework, the model is typically assigned via
the likelihood, i.e., the conditional distribution of data given unknown parameters, and a joint prior distribution for all the parameters, which represents belief \textit{prior} to seeing the current data. 
Inference is based on the posterior distribution of the parameters, i.e., the conditional distribution of parameters given data, which is proportional to the likelihood times the prior, as Bayes' theorem states. 
Except for simple models, it is not possible to analytically compute the posterior distribution and one usually resorts to numerical methods. 
A popular class of such methods, which enjoys several theoretical properties, is Markov chain Monte Carlo (MCMC) algorithms; see, for instance, \cite{brooks2011handbook}.  \cite{martin2020computing} is a recent review on Bayesian computation, where alternatives to MCMC such as variational inference or Laplace approximation are also discussed.
As the name says, an MCMC algorithm builds a Markov chain whose limiting distribution is the desired target, i.e., the posterior distribution in our case.
Consequently, integrals of interest are approximated via Monte Carlo integration.

One of the main drawbacks of Bayesian inference is that MCMC methods can be extremely demanding from the computational point of view. Moreover, the design of efficient MCMC algorithms and their practical implementation is not a trivial task, and thus might preclude the use of these methods to non-specialists.
Nonetheless, Bayesian statistics has greatly increased in popularity in recent years, thanks to the growth of computational power of computers and the development of several dedicated software products. In particular, these software programs are able to generate MCMC samples from the posterior in a black-box fashion
starting from the specification of a Bayesian model.
This software is commonly referred to as \emph{probabilistic programming languages} (PPLs). We can roughly classify the users of PPLs into three categories. The first one consists of applied scientists and non-statisticians in general, who can use Bayesian modeling to include their domain knowledge in the model; see for instance \cite{baydin2019efficient} for applications in the field of particle physics and \cite{do2019relativistic}, co-authored by the Nobel winner dr. Andrea Ghez, where the authors use Bayesian inference to validate their hypothesis. 
A second category consists of applied statisticians, whose focus is towards statistical modelling and real-world applications; they typically prefer to rely on the PPL software to compute posterior inference, as it is more practical than implementing an MCMC algorithm for the complex models they generally assume \citep[see, e.g.,][]{dutta2021joint, gramatica2021bayesian, gelman2021slamming}. 
Finally, methodological statisticians can use this software to compare newly proposed algorithms or models against state-of-the art methods \citep[e.g.,][]{rui2021fast, nemeth2021stochastic}. 
 Also we, the authors, belong to this last category. For instance, \cite{beraha2020spatially} provide a C++ code for the MCMC algorithm associated to their model, but they use Stan to compute posterior inference from a competitor model for which deriving the MCMC algorithm from scratch would have been too long.
 
In this paper, we consider JAGS \citep{Plummer}, Stan \citep{carpenter2017stan, stan2018stan} and NIMBLE (\citealp{NIMBLE}, \citealp{nimble_software}). 
At this moment, these platforms are the most popular probabilistic programming languages in the Bayesian community, as testified by  the number of citations received by those papers.
For instance, \cite{carpenter2017stan} has received 4076 citations by Google Scholar, of which 266 times in scientific papers in the area of statistics, probability and computer science (as reported by ISI Web of Science Core Collection). Similarly, \cite{Plummer} has received a total of 4597 citations according to Google Scholar (though we were not able to further analyze the sources), while \cite{NIMBLE} has been cited 236 times in total of which 35 times by papers in statistics, probability or computer science. 
Since they are based upon different MCMC algorithms, it is usually the case that, when considering the same Bayesian model, one of them is more \emph{efficient}, i.e., it is able to generate the samples in a smaller amount of time. 
Moreover, since the output of MCMC algorithms is a realization from a Markov chain, the elements of the chain, i.e. the samples, will be autocorrelated. 
As strongly autocorrelated samples provide less information, it is important to evaluate also the \emph{quality} of the MCMC samples produced by the different software programs.
 Given that simulations from the posterior of a Bayesian model can take, in some applications, several hours or even days, knowing which software platform is \textit{best suited} for the model before running any MCMC simulation is of great interest, since the wrong choice might lead to a huge loss of time.

Comparing and benchmarking probabilistic programming languages has been a hot topic for several years. The vast majority of these comparisons are available on the web in the form of blog posts or tutorials. While these resources are undoubtedly useful, they usually present analyses only on one particular case study or one particular class of models. As such,  getting a comprehensive overview of the pros and cons of each PPL  requires going through dozens blog posts. 
For instance, at the following link \url{https://nature.berkeley.edu/~pdevalpine/MCMC_comparisons/some_ARM_comparisons/election/nimble_election88_comparisons.html}, 
the authors consider a generalized linear model with random effects and compare the efficiency of MCMC sampling using Stan, JAGS and NIMBLE. 
They conclude that NIMBLE can be the most  \emph{efficient} software (with respect to a well defined criterion we do not report here)  after a fair amount of fine tuning, while Stan is the most efficient software when used with default parameters.  Another example of such blog posts is \cite{boelstad:2019a}, who presents a comparison between JAGS and Stan for linear models.
Both blog posts are useful and instructive for novices willing to implement their Bayesian models using a PPL. However, these posts offer a limited perspective, constrained to the particular classes of model considered, so that it is unclear if their  conclusions are likely to hold in other scenarios.

Given that many blog posts are written by newcomers to the field, it is often the case that the results reported are misleading or incomplete. Unknown authorship in blogs makes things even harder.
However, some blogs from the software authors themselves can be very helpful.
For instance, the blog post \cite{carpenterblog},  authored by one of the main contributors of the Stan software, points out several flaws that affect the results of the comparison, such as excessive thinning of the chains, the use of too many burn-in iterations and the study of simple models only. Our point here is not denying the usefulness of these webpages or blogs, but rather the acknowledgement that this type of comparison should be published in the form of a scientific article.

To the best of our knowledge, only \cite{monnan2017faster} organize a systematic review of PPLs in the form of a paper. However, their audience is made of ecologists rather than statisticians and, as such, only models for ecological applications are considered for the comparison. Their focus is on mixed-effects models and state-space models and NIMBLE is not included in the comparison since, at that time, it was at an embryonic stage. Moreover, they do not examine the effect of the prior nor the effect of the sample size or of the dimension of the parameter space in their comparison.
Though they concentrate on the underlying algorithms, \cite{betancourt2015hamiltonian} propose a comparison between several MCMC algorithms on simulated datasets in highly-dimensional settings. \cite{betancourt2015hamiltonian} also point out that, most of the times, comparing the mixing of the MCMC chains is not enough and that one must separately check that the MCMC chains given by different software platforms produce consistent estimates. This point is often overlooked in blog posts, while we consider it as well (see Section~\ref{sec:comparison}).

In this review, we carry out a systematic comparison of  JAGS, Stan and NIMBLE (here listed in chronological order of their release), on a wide class of models, prior distributions, and data generating mechanisms. 
The final goal is to present a detailed analysis of the strengths and weaknesses of these popular PPLs to statisticians or applied scientists, in one of the three categories of audience listed above. In this way, we wish to contribute to make them fully aware of the pros and cons of this software. 
 Codes for all the models considered in this paper are publicly available at \url{https://github.com/daniele-falco/software_comparison}, and new users can start developing their code from our optimized examples.
Specifically, we consider the classes of linear regression, logistic regression, accelerated failure time and mixture models. 
For each class of models, we then consider several alternative prior distributions among those typically adopted. In total we have tested nine Bayesian models.
A larger comparison can be found in \cite{falcothesis} where over 30 models have been considered.
All the analyses are carried out on synthetic datasets, which we obtain by simulating data from the likelihood after having fixed the values of all the unknown parameters, and, when needed, of covariates.
Throughout our analyses we vary the dimension of the datasets, i.e. considering different sample sizes, different number of parameters and, when needed, different number of groups.
For some models, we  have also tested the ability of these PPLs to generate several independent MCMC chains in parallel, with a special focus on the amount of memory required.
Despite the huge number of cases analysed, obtained considering different models, priors, dimensionalities of the datasets and number of chains, it  is impossible to make the comparison through 
 all the models that a user could contemplate. 
 By interpreting the findings of our comparison through the knowledge of the MCMC algorithms adopted by the different software programs, we are allowed to draw \textit{larger} conclusions. 

Summing up,
we believe that this review article can provide general guidance to non-specialist and specialist audience, supporting their choice of which PPL to use even when considering models that are not analysed here.
All simulations were performed using an ASUS LAPTOP-FHEVTGN6 with processor Amd Ryzen 7 3750H, RAM 16GB.

The remainder paper is organized as follows. 
Section~\ref{sec:software} introduces basic notions on MCMC algorithms and the main features of the PPLs under comparison. We also introduce some statistics to compare the software programs. Section~\ref{sec:models} presents all the models we tested and it explains the data generating process for the examples. The findings of our comparison are in Section~\ref{sec:comparison}. The article concludes with a discussion in Section~\ref{sec:discussion}. Appendix~\ref{sec:details} contains details on monitoring the convergence of the simulated Markov chains, while 
the glossary 
of all probability distributions in the article is listed in Appendix~\ref{sec:appendix_distr}. Appendix~\ref{sec:appendix_gof} reports explicit formulas of the goodness-of-fit indexes we show in this article. 
Appendix~\ref{sec:repeated_simulations} illustrates the comparison  among the software through statistics averaged over different simulated datasets to ensure robustness of  the conclusions.

\section{Software and algorithms under comparison}
\label{sec:software}

In  this section, we introduce the main features of the software platforms we  compare and the description of the relative MCMC algorithms. We also describe  here the procedure we have applied to monitor and compare the relative MCMC samples.  See also  \cite{martin2020computing}, Section~5.

\subsection{A primer on Markov chain Monte Carlo}

Before giving details on the different software platforms, we shortly review MCMC algorithms.
We consider data $\bm y = (y_1, \ldots, y_n)$ with $y_i \in \mathbb{Y} \subseteq \mathbb{R}^d$. Given the likelihood $\mathcal{L}(\bm y \mid \bm \theta)$, i.e., the (conditional) joint distribution of $\bm y$ given parameter $\theta \in \Theta$ (e.g. $\Theta \subseteq \mathbb{R}^p$),
and the prior density $\pi(\bm \theta)$, the posterior density is derived by Bayes' theorem as
\[
    \pi(\bm \theta \mid \bm y) = \frac{\mathcal{L}(\bm y \mid \bm \theta) \pi(\bm \theta)}{\int_\Theta \mathcal{L}(\bm y \mid \bm \theta) \pi(\bm \theta) \mathrm{d}  \bm \theta} \propto \mathcal{L}(\bm y \mid \bm \theta) \pi(\bm \theta),
\]
where the denominator is the marginal distribution of $\bm y$ and, in general, is not available in closed form.

In Bayesian inference, MCMC methods are used to obtain samples $\bm \theta^{(1)}, \bm \theta^{(2)}, \ldots$ from $\pi(\bm \theta \mid \bm y)$ when the posterior density is known only up to a normalizing constant. See \cite{brooks2011handbook} for a detailed review. Under mild assumptions on the transition kernels, the theory of general state space Markov chains guarantees that the limiting and the stationary distributions coincide, so that, for $n$ large enough, the marginal distribution of $\bm \theta^{(n)}$ (and of all subsequent $\bm \theta^{(n+1)},\bm \theta^{(n+2)},\ldots$) is approximately $\pi(\bm \theta \mid \bm y)$ (i.e., the chain has reached stationarity). Note that, however, $\bm \theta^{(n)}, \bm \theta^{(n+1)}, \ldots$ are not independent.

The cornerstone in MCMCs is the Metropolis-Hastings (MH) algorithm \citep{Metropolis, Hastings}, where, given the current state of the chain $\bm \theta$, a new value $\bm \theta^\prime$ is proposed from a density $q(\bm \theta^\prime \mid \bm \theta)$ and accepted with probability $\min(1, \alpha)$, where $\alpha$ is the acceptance ratio
\[
    \alpha := \frac{\pi(\bm \theta^\prime \mid \bm y) q(\bm \theta \mid \bm \theta^\prime)}{\pi(\bm \theta \mid \bm y) q(\bm \theta^\prime \mid \bm \theta)} = 
    \frac{ \mathcal{L}(\bm y \mid \bm \theta^\prime) \pi(\bm \theta^\prime) q(\bm \theta \mid \bm \theta^\prime)}{\mathcal{L}(\bm y \mid \bm \theta) \pi(\bm \theta) q(\bm \theta^\prime \mid \bm \theta)}.
\]
Note that the intractable normalizing constants cancel out from the numerator and the denominator.
Several popular MCMC algorithms are special cases of the MH algorithm, for different choices of the proposal distribution $q(\bm \theta^\prime \mid \bm \theta)$.
Although the ergodic theory of Markov chains ensures convergence to the limiting distribution for most (reasonable) choices of $q(\cdot \mid \cdot)$, the efficiency of the MCMC algorithm can be extremely sensitive to the specific choice.

A  simple example is the Random-Walk Metropolis Hastings algorithm, where $q(\bm \theta^\prime \mid \bm \theta) = \mathcal{N}(\bm \theta, \sigma^2 I)$, which is straightforward to code. 
\MBtext{In} order to get non-zero acceptance rates, $\sigma^2$ must be small (especially when the parameter space is highly dimensional), and consequently the chain moves very little at each iteration.
This has two side-effects. First, in order to reach the limiting distribution, several hundreds of thousands iterations might be required. Second, the chain usually has a \emph{poor mixing}, i.e., it explores the support of the posterior very slowly, so that the samples are highly autocorrelated. For further details on autocorrelation and its (negative) impact on inference see Appendix~\ref{sec:details}.  
Improvements of the Random-Walk MH exploit the geometric structure of the posterior distribution to design the proposal distribution. 
A particularly popular algorithm is Hamiltonian Monte Carlo \citep[HMC,][]{hmc, hmc2, neal2012}, where the gradient of the posterior density is used when proposing a new value. This allows HMC to propose (and accept) new values of the parameters that are far away from the current state, unlike the traditional Random-Walk MH. Since the computation of the gradient is mandatory, only continuous parameters can be handled using HMC.

Another particular case of the MH algorithm is the Gibbs sampler \citep{Geman}, where the vector $\bm \theta$ gets updated one component at a time, by sampling the $j$-th component $\theta_j$ from its \emph{full conditional}, that is the conditional law of $\theta_j$ conditioned on data and all the other parameters.  In the \emph{blocked} Gibbs sampler, 
a block of multiple parameters can be updated in a single step, always sampling from their full conditional distribution. 
Gibbs sampling is convenient when the posterior of $\bm \theta$ is intractable but full conditional of the $\theta_j$'s are known and easy to sample from.
As shown in \cite{lunn2000}, if the full conditional distribution of some $\theta_j$ is itself intractable, the new value for $\theta_j$ can be sampled using a single step of any MCMC algorithm.

\subsection{The software platforms}
	
JAGS is the acronym of \textit{Just Another Gibbs Sampler}. It is written in  C++ and interfaced with the R language via \texttt{rjags} \citep{rjags} or via \texttt{runjags} \citep{runjags}. 
JAGS relies on Gibbs sampling to update each block of parameters.
If one of the full conditionals is intractable, when this distribution is log-concave, the sampling is obtained through adaptive rejection sampling \citep{Gilks}, otherwise one step of slice sampling \citep{Neal} or Metropolis-Hastings algorithm  is performed.

Stan is an open source program written in C++, whose R interface is provided by the package \texttt{rstan} \citep{rstan}. Stan implements the Hamiltonian Monte Carlo algorithm and its variant  No-U-Turn Sampler, NUTS for short \citep{nuts}. NUTS is more convenient than basic HMC since it is able to automatically tune hyper-parameters  of the algorithm. 
Note that Stan inherits the unfeasibility of discrete parameters  from HMC.
	
NIMBLE \citep{nimble_software}, which stands for  \textit{Numerical Inference for statistical Models for Bayesian and Likelihood Estimation},  was first motivated as a software to simulate from the posterior of hierarchical models, but it can also be used for other models. NIMBLE may rely on different MCMC algorithms, and the specific algorithm is chosen according to the characteristics of the likelihood and the prior specification. In general, NIMBLE uses the Random-Walk or block Random-Walk Metropolis-Hastings algorithm, but in some cases, when the full conditional distribution is available in analytic form, it resorts to Gibbs sampling. Moreover, NIMBLE may use different MCMC algorithms for different blocks of parameters.

\subsection{Comparing MCMC chains and software}
\label{sec:indexes}

To compare  the software, we first need to fix a proper metric. In the literature, many statistics have been introduced to verify the information provided by the output of a MCMC algorithm, and their computation is usually automatically performed by some packages such as, for example, \texttt{CODA} \citep{coda_p}. We have adapted some of those heuristics to fulfil our goal.
Specifically, we monitor separately the \textit{quality} of the chains and the \textit{efficiency} of the software, as discussed below. See also the blog post \url{https://nature.berkeley.edu/~pdevalpine/MCMC_comparisons/nimble_MCMC_comparisons.html} for some general guidelines to compare different PPLs.

All the MCMC chains have been run for a finite number of iterations. We discard the first number $N_{b}$ of iterations because we have not reached stationarity yet, or store less iterations after burn-in, i.e.,  only one every $N_{thin}$ iterations  is saved (this is the thinning  mentioned in the Introduction).
A key quantity that is usually monitored is the effective sample size ($ess$). Informally, for a sample of size $N_s$ from Markov chain, $ess$ can be interpreted as the number of independent and identically distributed (iid) draws that contain the same amount of information of the whole chain; see Appendix~\ref{sec:details} for further detail. For each unidimensional parameter $\theta_j$, we consider $\mathcal{E}_j := ess / N_s$, i.e., the fraction of \virgolette{iid samples} contained in the chain, and then report the average $\mathcal{E}$ of the $\mathcal{E}_j$'s across all the parameters, as a measure of the \emph{quality} of the chains.
Assuming that, after the burn-in   phase, the chains have reached stationarity, the ratio $\mathcal{E}_j$ does not depend on the length of the chain and it represents the quality of the sampling, since it is small in presence of high autocorrelation and close to one in the opposite case. 
In some cases, it might happen that $ess > N_s$; chains where this phenomenon occurs are called \emph{antithetic}. In order to have $\mathcal{E} \in [0, 1]$, we always assume $N_{thin}$ equal to 2, that is we discard one every two iterations of the Markov chain, thus avoiding the antithetic behavior.

To measure the \emph{efficiency} of the software, we focus on runtimes. 
Although we ran our simulations in R, which is an interpreted programming language, all the three PPLs require a compilation phase  
before executing the actual MCMC simulation, because they rely on efficient implementations in C/C++ underneath.
We considered separately the compilation time ($t_c$) and the sampling time ($t_s$). The efficiency of the sampling is monitored by the ratio $N_{it}$/$t_s$, where $N_{it}$ is the total number of iterations (including burn-in). This ratio measures how fast the sampling is performed.

Usually, the $ess$ and the sampling time $t_s$ are combined into $\mathcal{E} / t_s$, i.e. the effective sample size per second. This index has the advantage of providing one single number to compare the software programs. However, for JAGS, NIMBLE and Stan, the amount of time that the sampling must be run for cannot be fixed, so that using $\mathcal{E} / t_s$  does not give an estimate of the final effective sample size.  Similarly, this software cannot fix
 the minimum $ess$ that must be reached before stopping the simulation of the Markov chain, so that  $\mathcal{E} / t_s$ is not an estimate of the runtime required. Instead, in JAGS, NIMBLE and Stan, users must specify the number of burn-in (or adaptation) iterations and can control the final sample size. Hence, our recommendation is to report $\mathcal{E}$ and $t_s$ separately.  

Finally, as suggested in \cite{betancourt2015hamiltonian}, we check that we get consistent estimates from every platform-specific MCMC chain by monitoring posterior predictive goodness-of-fit indexes, such as the log pseudo marginal likelihood (LPML),
the Watanabe-Akaike information criterion (WAIC), 
and the Kullback--Leibler divergence between the  true distribution generating the data and the posterior predictive distribution obtained from the MCMC. We also compute the difference between the posterior mean of the parameters and their  \virgolette{true value} (i.e., the value used to simulate the data) for all the platforms.  See Appendix~\ref{sec:appendix_gof} for their definition.

\section{Models and data}
\label{sec:models}

This section describes the Bayesian models that we consider for our comparison and gives further details on the data generating process.
Appendix~\ref{sec:appendix_distr} reports all the probability distributions considered here, together with the notation adopted.

\subsection{The Bayesian models}
\label{sec:bay_mod}

We consider four classes of Bayesian models, i.e. linear models (LMs), logistic regression models (LRs), mixture models (MMs) and accelerated failure time models (AFTs). Below, for each of these models, we  introduce the likelihood and prior distributions we consider in Section~\ref{sec:comparison}. Table~\ref{tab:hyperp} reports values of the hyperparameters in the priors, unless otherwise stated.

\begin{table}[tbp]
	\centering 
	\begin{tabular}{c | c }\hline 
		\eqref{eq:lm_conj}&$\sigma_{0}^2$=1, $\eta_{0}$=$10^{-4}$\\
		\eqref{eq:lm_wi}&$M$=100, $d_0$=2.5\\
		\eqref{eq:lm_ni}&$M$=100, $\sigma_{0}$=1000\\
		\eqref{eq:lm_lasso}&$\lambda_0$=0.1, $\nu_0$=$10^{-4}$, $\sigma_{0}^2$=1\\\hline
		\eqref{eq:lr_normal}&$b_{0}^{2}$=10\\
		\eqref{eq:lr_lasso}&$\lambda_0$=0.1\\\hline
		\eqref{eq:mix_prior}&$a_0$=1, $b_0$=1, $c_0$=1, $d_0$=1\\\hline
		\eqref{eq:aft_nh}&$b_{0}^{2}$=10, $\lambda_0$=1\\
		\eqref{eq:aft_ni}&$M$=100, $\sigma_{0}$=1000\\\hline
	\end{tabular}
	\caption{Values of the hyperparameters.}
	\label{tab:hyperp}
\end{table}

\paragraph{Linear models}

For data $\{(y_i, \bm x_i), i=1,\ldots,n\}$ such that $y_i \in \mathbb{R}$ and $\bm x_i \in \mathbb{R}^p$ for all $i$,  a linear model assumes the likelihood
\begin{equation}
		y_{i}|\bm{\beta}, \sigma^2, \bm{x}_{i} \ind  \mathcal{N}(\bm{x}_{i}^{T}\bm{\beta}, \sigma^2)\hspace{10pt}  i=1,\dots,n
	\label{eq:lm_lik}
\end{equation}
where $\bm \beta \in \mathbb{R}^p$ is the unknown vector of regression coefficients and $\sigma^2>0$ is the variance (independent of $\bm x_i$'s).
We consider four   prior distributions:
\begin{alignat}{3}
\bm \beta |\sigma^2 \  &\sim  \ \mathcal{N}_{p}(\bm{0}, \sigma^{2}I),	\ &  &\ &\sigma^2 \sim & \ \mathcal{IG}(\eta_0/2, \eta_0 \sigma_{0}^2/2) \tag{LM-C}	\label{eq:lm_conj} \\
 \beta_{j}\  &\iid \ \mathcal{N}(0,M^2) \ &j=1,\dots,p, &	 & \sigma \sim & \ \mathcal{HC}(0,d_0)  \tag{LM-WI} \label{eq:lm_wi} \\
		\beta_{j}\  &\iid \ \mathcal{N}(0,M^2) \ &j=1,\dots,p, &	 & \sigma \sim & \ \mathcal{U}(0,\sigma_0)  \tag{LM-NI} \label{eq:lm_ni}\\
 \beta_{j} |\lambda^2  &\iid  \mathcal{DE}  \left(0,  1 / \sqrt{\lambda^2} \right) \ &j=1,\dots,p, & \quad
		& \lambda^2 \sim \ &  \mathcal{E}(\lambda_0), \ \sigma^{2}  \sim \mathcal{IG}\left(\frac{\nu_{0}}{2},\frac{\nu_{0}\sigma_{0}^{2}}{2}\right) \tag{LM-L} \label{eq:lm_lasso}	
\end{alignat}
See Table~\ref{tab:distributions} for notation.
Prior \eqref{eq:lm_conj} is the conjugate prior that can be found in most textbooks on Bayesian statistics; see, for instance, \cite{jackman2009bayesian}. 
The posterior distribution under \eqref{eq:lm_conj} belongs to the same parametric family of the prior and is available in closed form; hence, MCMC for this model is not necessary, but we include it as a \virgolette{sanity} check.
Priors \eqref{eq:lm_wi} and \eqref{eq:lm_ni} were proposed in \cite{gelman2006prior} as priors for the variance parameters in hierarchical models, but they are very often assumed as priors for any variance parameter. In particular, following \cite{gelman2006prior}, 
\eqref{eq:lm_wi} is named as \emph{weakly-informative} prior, while 
\eqref{eq:lm_ni} is called \emph{non-informative} prior.
Prior \eqref{eq:lm_lasso} introduced in \cite{park2008bayesian} with the name of Bayesian lasso, is a popular \virgolette{shrinkage} prior, as the double exponential distribution assigns significant mass to values near to zero. This prior is commonly used for  covariate selection problems.

\paragraph{Logistic regression models}
Logistic regression is a particular case of generalized linear models (GLMs), which extend the linear regression model in \eqref{eq:lm_lik} to account for non-continuous or non-Gaussian responses $y_i$. In particular, for binary responses $y_i \in \{0, 1\}$, with $\bm x_i\in\R^p$ as before, we assume the likelihood 
\begin{equation}
		y_i|\bm{\beta},\bm{x}_{i} \ind \text{Be} \left(\frac{1}{1+\e^{-\bm{x}_{i}^{T}\bm{\beta}}}\right) \hspace{20pt} i=1,\dots,n.
		\label{eq:lr_lik}
\end{equation}
Note that the parameter of the Bernoulli distribution as in \eqref{eq:lr_lik} corresponds to the logit  model. 
As before, we compare more than one prior distribution for $\bm \beta$, to understand the effect of the prior on the MCMC efficiency. In particular, we assume one of the following two priors
\begin{alignat}{2}
\beta_{j} & \iid \ \mathcal{N}(0,b_{0}^{2})\hspace{20pt} &j=1,\dots,p, &\tag{LR-N} \label{eq:lr_normal} \\
\beta_{j} |\lambda^2 & \iid \mathcal{DE}(0,  1/\sqrt{\lambda^2}) \hspace{10pt} &j=1,\dots,p, & \qquad
		\lambda^2\  {\sim} \ \mathcal{E}(\lambda_0) \tag{LR-L} \label{eq:lr_lasso}.
\end{alignat}
Under prior \eqref{eq:lr_normal} the $\beta_j$'s are a priori independent and normally distributed. This prior is advocated in \cite{chopin} as \textit{\virgolette{... a proper prior that assigns a low probability that the marginal effect of one predictor is outside a reasonable range}}.
Prior \eqref{eq:lr_lasso} is the same as the lasso prior for the linear model.

\paragraph{Mixture models}

Mixture models are a popular framework for density estimation and model-based clustering. See \cite{mixture_book} for a recent review. We assume here a finite mixture model of univariate normal densities as 
\begin{equation}
		y_i| \bm \mu, \bm \sigma^2, \bm{p} \iid \sum_{h=1}^{H}p_h \mathcal N (\cdot|\mu_h, \sigma^2_h) \quad i=1,\dots,n \label{eq:mix_lik}
\end{equation}
with $H$ fixed to an integer value. Here $\bm \mu=(\mu_1,\ldots,\mu_H)$, $\bm \sigma^2=(\sigma_1^2,\ldots,\sigma_H^2)$ and  $\bm p=(p_1,\ldots,p_H)$  where $p_h$ represents the weight associated to the $h$-th \emph{component} of the mixture \eqref{eq:mix_lik} with $p_h > 0$ for each $h$ and $\sum_h p_h = 1$.
A common strategy to perform posterior simulation via MCMC consists in introducing latent variables $z_i \in \{1, 2, \ldots, H\}$ for each observation $i$, and expressing \eqref{eq:mix_lik} as
\begin{equation}\label{eq:mix_lix_lat}
\begin{aligned}
    y_{i} | z_{i}, \bm \mu, \bm \sigma^{2} & \ind \ \mathcal{N}(  \mu_{z_{i}}, \sigma^{2}_{z_{i}})\quad i=1,\dots,n \\
    z_i | \bm p & \iid \text{cat}(H;\bm p)
\end{aligned}
\end{equation}

Stan allows only parametrization \eqref{eq:mix_lik} because of unfeasibility of discrete parameters, while JAGS and NIMBLE allow also \eqref{eq:mix_lix_lat}. 
In each software platform $H$ must be fixed to a positive integer value, though 
NIMBLE allows $H = +\infty$ when assuming a Dirichlet process prior \citep[see][for a review]{muller2015bayesian}, which might be an appealing feature in cases when finding the \textit{best value} for $H$ is  computationally demanding.

A priori we assume that 
\begin{equation}
\begin{aligned}
\mu_h |v^2 & \iid  \mathcal{N}( 0, v^2)  \ \ h = 1,\dots,H, \quad
v^2 & \sim \mathcal{IG}(a_0, b_0) \\
\sigma^2_h & \iid \mathcal{IG}( c_0, d_0) \ \ h = 1,\dots,H\\
	\bm{p} & \sim \mathcal{D} (1,\dots,1).
\end{aligned}
\tag{MM} \label{eq:mix_prior}
\end{equation}
Prior \eqref{eq:mix_prior} is equivalent to assume $\mu_h$ and $\sigma_h$ independent  for any $h$ and $\mu_h$ marginally distributed according to a  $t$-density. Moreover  the common parameter $v^2$ induces prior exchangeability of $\mu_1,\ldots,\mu_H$.
The posterior distribution of a mixture model has a particularly complex geometry due to the so-called \virgolette{label switching}, i.e., the likelihood \eqref{eq:mix_lik} is invariant under any permutation of the indexes $h$. The  joint posterior distribution of $\bm p$ and $\bm \mu, \bm \sigma^2$ is multi-modal, with each mode corresponding to a different labelling of the components. It is often the case that even tailored MCMC algorithms cannot explore properly the posterior and get stuck on one mode of high density \citep[see][]{celeux2019computational}.  The common parameter $v^2$ has been introduced to monitor convergence across the platforms without label switching issues. 
All in all,  mixture models offer a challenging benchmark for PPLs.
	
\paragraph{Accelerated failure time models}

AFT models are useful when dealing with right censored data; see \cite{BIDA} for a short review.
We observe data $(y_i, \delta_{i}, \bm x_i)$ for $i=1,\ldots,n$, with $y_{i} = \min\{T_{i}, C_{i}\}$ and $\delta_{i}=1$ if $T_{i}\leq C_{i}$ and $\delta_{i}=0$ otherwise.  The random variables $T_i$ and $C_i$ represent failure time and censoring time, respectively. As before, covariates $\bm x_i\in\R^p$. We assume that, conditioning to regression parameters $\bm\beta$ ($p$-dimensional) and parameter $\sigma>0$, data $T_i$ are independently distributed according to the following regression model in the log-scale: 
\begin{equation}
\log(T_{i})=\bm{x}_{i}^{T}\bm{\beta}+\sigma\epsilon_{i}  \hspace{20pt} i=1,\dots,n
\label{eq:AFT_logscale}
\end{equation}
where the $\epsilon_i$'s represent iid \virgolette{errors} with cumulative distribution function $F_{\epsilon}$, i.e.,
\[
    \epsilon_{i}\stackrel{iid}{\sim} F_{\epsilon}(u)=1 -\exp \left(-(\log2)\e^{u} \right) \hspace{20pt} u \in \mathbb{R}
\] 
The factor $\log 2$ in the expression of $F_{\epsilon}$ above makes its median equal to 0. 
Equivalently, we have 
\begin{equation}
		T_{i}| \bm\beta, \sigma, \bm{x}_{i}  \ind \text{Wei}\left(\frac{1}{\sigma}, (\log 2) \e^{-(\bm{x}_{i}^{T}\bm{\beta})/\sigma}  \right) \quad i=1, \ldots, n
		\label{eq:aft_2}
\end{equation}
where Wei denotes the Weibull distribution according the parameterization as in  Appendix~\ref{sec:appendix_distr}, Table~\ref{tab:distributions}. As
it is standard in this context, we assume that $T_i$ and $C_i$ are (conditionally) independent and that the distribution of $C_i$ does not depend on the parameters of interest $\bm\beta$ and $\sigma$ (non-informative censoring assumption).

As a prior for $\bm \beta$, we could assume any prior distributions considered above for linear models since \eqref{eq:aft_2} is equivalent to a linear regression model in the log-scale, see \eqref{eq:AFT_logscale}. Here we consider only two prior distributions:
\begin{alignat}{2}
	\beta_{j} & \iid \mathcal{N}(0, b_{0}^{2}) \enspace &j=1,\dots,p, & \qquad
	\sigma \sim \ \mathcal{E}(\lambda_0), \tag{AFT-NH} \label{eq:aft_nh} \\
	\beta_{j} & \iid \mathcal{N}(0,M^2) \enspace &j=1,\dots,p, & \qquad
	\sigma\  {\sim} \ \mathcal{U}(0,\sigma_0) \tag{AFT-NI} \label{eq:aft_ni}.
\end{alignat}
We refer to  the first prior as  \emph{non-hierarchical} (AFT-NH),  while the second one, (AFT-NI), is the same non-informative prior (LM-NI) we have considered for linear models.

\subsection{Data generating process}
\label{subse:simulation_data}

For each model, we have generated synthetic datasets by simulating from the likelihood, for fixed values of the parameters, and, when needed, of the covariates. 

Specifically, the values of regression parameters, i.e. the components of parameters $\bm{\beta}$ in LMs, LRs and AFTs, have been fixed equal to a random value sampled uniformly between -1 and 1 for AFT and, for the other models, between -7 and 7. The value of $\sigma^2$ in LMs and AFTs has been randomly sampled between 2 and 10. 
When considering the lasso prior for LMs and LRs (see \eqref{eq:lm_lasso} and \eqref{eq:lr_lasso}), we have fixed the value of some components $\beta_j$ equal to zero to test also the effectiveness of the variable selection prior. 
In particular, considering $p=30$ and $p=100$ covariates, we have studied three different settings. First, we set two regression parameters (out of $p$) equal to zero, then we set half regression parameters equal to zero, and lastly we set all the regression parameters (but two) equal to zero.

Covariates have been always simulated from independent standard normal distribution and the first element of each $\bm x_i$ has been fixed to 1 to include the intercept term.
LMs under prior \eqref{eq:lm_conj} have also been analysed considering, instead of continuous covariates, binary covariates independently simulated from a Bernoulli distribution. When $p=4$, the hyperparameter of the Bernoulli distribution was fixed equal to 0.1, 0.5 and 0.8 for the second, third and fourth covariate values in $\bm x_i$, respectively, for each $i$.
When $p=16$, for each $i$,  we have simulated each of the 15 covariates values in $\bm x_i$ from the Bernoulli distribution with hyperparameters
$0.1,0.2,0.3,0.4,0.5,0.5,0.5,0.5,0.5,0.5,0.5,0.6,0.7,0.8,0.9$, respectively. When $p$=50, the 49 values in $\bm x_i$ have been independently sampled from the Bernoulli distribution with parameter randomly sampled between 0.05 and 0.95.
	
For  MMs, we have considered $H=2$ and $H=4$ components. In the data simulation process, each weight $p_h$ was set equal to $1/H$. When $H=4$, we fixed $\bm{\mu}=(-4,0,2,6)$, while when $H=2$, $\mu_1$ was sampled uniformly between -2 and 0 and $\mu_2$ between 1 and 3. The standard deviations $\sigma_{j}$ were all fixed equal to 1.	
	
In the case of AFT models, denoting by $100k\%$ the percentage of right-censored data to simulate,
 we have assumed three scenarios,  with $k=0.2, 0.5$ and $0.8$, respectively. For any $i$, we have simulated first the failure time $T_i$ from \eqref{eq:aft_2}, after having fixed $\bm{x_i}$, $\bm{\beta}$ and $\sigma$ as above.
As suggested by \cite{survivaldata}, we have simulated the censoring time $C_i$ from the Weibull distribution with parameters $1/\sigma$ and $\frac{k}{1-k} (\log 2) \e^{-(\bm x_i^T \bm \beta)/\sigma} $.
Under this choice, on average, $100k\%$ of data result right-censored; see \cite{survivaldata}. 
 
Comparison across the three software platforms has always been made using the same simulated dataset, though we test different datasets for each model.
	

\section{Software comparison on simulated datasets}
\label{sec:comparison}

Here we present our findings. Specifically, Section~\ref{sec:software_comparison} shows the comparison for all the models (and prior distributions) presented in Section~\ref{sec:models}. We vary the dimension of the datasets, i.e. considering different sample sizes, different number of parameters and, when needed, different number of groups.
To conclude that our interpretation is robust, we compute average values of the statistics under different simulated datasets (20, 30 or 50 datasets, according to the example); see Appendix~\ref{sec:repeated_simulations}.
In Section~\ref{sec:parallel_chains} we discuss the efficiency of parallelization in the three software platforms.
	
\subsection{Software comparison}
\label{sec:software_comparison}

The code that implements each model in JAGS, Stan and NIMBLE (through syntax-specific text files) 
can be found at \url{https://github.com/daniele-falco/software_comparison}.  The PPLs have been run through their R interface \citep{R}.
Results presented in this section are obtained running only one single MCMC chain for each model. The models marked with * in Tables~\ref{tab:lm_res}-\ref{tab:aft_res} have been further analysed in Section~\ref{sec:parallel_chains} where we test our conclusions by running several parallel chains.
Moreover, Appendix~\ref{sec:appendix_gof}  compares the platforms via
a predictive goodness-of-fit index (LPML  or WAIC) and the error between the posterior mean of parameters and the true value used to generate the data. For mixture models, we also computed the Kullback--Leibler divergence between the data generating distribution and the predictive distribution.

All the three software platforms are based on efficient C++ implementations and provide a user-friendly interface via R. JAGS and NIMBLE's interface are based on the BUGS language, while Stan has developed its own language.
Once the model has been declared (i.e., written in a text file) using either BUGS or Stan language, a compilation phase must  follow.
In JAGS, this should be better called a \virgolette{transpilation} phase: a compiler is invoked to translate from the BUGS language and creates the executable file by linking it to the JAGS library.
Stan relies on C++ \virgolette{templates}, which, in short, allow faster runtime performance but require longer compilation times. In fact, the Stan library must be compiled together with the model and not just linked to it.
Also NIMBLE relies on the compilation of the model in C++, but it does not rely on C++ templates, so that fewer lines of code must be compiled each time.
In our examples, the compilation phase for Stan required from 110  to 150  seconds approximately, while for NIMBLE we experienced much more variability: some models required as little as 40  seconds while others 250 seconds. For JAGS, the compilation is almost immediate and we do not report statistics related to it in our analysis.
Once compiled, Stan models can be saved and reloaded so that one can compile the model only once, while JAGS and NIMBLE require a new compilation/linking at each invocation. 

\paragraph{Linear model}

\begin{table}[htbp]
		\centering 
		\caption*{\Large{Linear Model}}
			\begin{tabular}{c | cc | cc | ccc |ccc}\hline
				& & &\multicolumn{2}{c|}{\textbf{JAGS}} &\multicolumn{3}{c|}{\textbf{STAN}} &\multicolumn{3}{c}{\textbf{NIMBLE}} \\
				& n & p & $\mathcal{E}_{\bm \beta}$ & $N_{it} / t_s$ & $t_c$ & $\mathcal{E}_{\bm \beta}$ & $N_{it} / t_s$  & $t_c$ & $\mathcal{E}_{\bm \beta}$ &$N_{it} / t_s$	\\ \hline
 				\parbox[t]{5mm}{\multirow{6}{*}{\rotatebox[origin=c]{90}{\eqref{eq:lm_conj}}}}
				& 100 & 4 & \textbf{100\%} & \textbf{3,667} & 147&96\%&1,833&86&14\%&11,000 \\
				& 1000&4  & \textbf{100\%} & \textbf{478} &147&99\%&157&85&14\%&1,571 \\
				& 100&16 & \textbf{99\%}& \textbf{1,571} & 147&84\%&611&81&4\%&5,500 \\
				& 1000*&16*& \textbf{100\%} & \textbf{200} &147&70\%&116&102&3\%&647\\
				& 2000*&30*& \textbf{99\%} & \textbf{58} &147&70\%&45&181&1\%&136\\
				& 30*&50*& \textbf{100\%} & \textbf{177} &147&93\%&42&92&1\%&2,750\\ \hline
				\parbox[t]{5mm}{\multirow{5}{*}{\rotatebox[origin=c]{90}{\eqref{eq:lm_conj} - Bin}}}
				& 100&4&  \textbf{100\%}& \textbf{3,667} &147&90\%&611&94&12\%&11,000\\
				& 1000&4& \textbf{95\%} & \textbf{550} &147&99\%&96&107&14\%&1,571\\
				& 100&16& \textbf{99\%} & \textbf{1,833} &147&86\%&244&100&3\%&5,500\\
				& 1000&16& \textbf{98\%} & \textbf{423} &147&70\%&36&123&3\%&647\\
				& 30&50&  \textbf{100\%} & \textbf{186} &147&90\%&26&107&1\%&2,750 \\ \hline
				\parbox[t]{5mm}{\multirow{4}{*}{\rotatebox[origin=c]{90}{\eqref{eq:lm_wi}}}}
				& 100&4&  \textbf{93\%}& \textbf{5,000} &135&97\%&1,920&49&39\%&294\\
				& 1000&4&  \textbf{100\%} & \textbf{357} & 135&99\%&130&60&45\%&176\\
				& 100&16&  \textbf{95\%} & \textbf{1,875} & 135&85\%&1,450&52&33\%&250\\
				& 1000*&16*&  \textbf{100\%} & \textbf{89} &135& \textbf{71\%} & \textbf{95} &69&42\%&93 \\ \hline
				\parbox[t]{5mm}{\multirow{4}{*}{\rotatebox[origin=c]{90}{\eqref{eq:lm_ni}}}}
				& 100&4& \textbf{97\%} & \textbf{5,000} &130&97\%&2,143&51&40\%&7,500\\
				& 1000&4& \textbf{100\%} & \textbf{455} &130&99\%&115&48&42\%&750\\
				& 100&16& \textbf{96\%} & \textbf{2,143} &130&84\%&1,250&46&33\%&1,875\\
				& 1000&16& \textbf{100\%} & \textbf{99} & 130 & \textbf{70\%} & \textbf{133} &73&41\%&163\\
				& 30&50&2\%&1,667&130& \textbf{78\%} & \textbf{205} &45&1\%&1,500
				\\\hline 
				\parbox[t]{5mm}{\multirow{8}{*}{\rotatebox[origin=c]{90}{\eqref{eq:lm_lasso}}}}
				& 100&$16^{(0)}$&74\%&323&138& \textbf{95\%} & \textbf{1,111} &67&35\%&1,667\\
				& 1000*&$16^{*(0)}$&89\%&20&138&\textbf{97\%} & \textbf{148} &79&43\%&133\\
				& 1000&$30^{(2)}$&-&-&138&\textbf{89\%} & \textbf{105} &104&41\%&70\\
				& 1000*&$30^{*(15)}$&87\%&10&138&\textbf{90\%} & \textbf{108} & 95&41\%&66\\
				& 1000&$30^{(28)}$&-&-&138&\textbf{90\%} & \textbf{78} &106&41\%&65\\
				& 1000&$100^{(2)}$&-&-&138&\textbf{97\%} & \textbf{17} &254&36\%&16\\
				& 1000&$100^{(50)}$&-&-&138&\textbf{99\%} & \textbf{29} &215&37\%&17\\
				& 1000&$100^{(98)}$&-&-&138&\textbf{92\%} & \textbf{19} &216&38\%&17
				\\\hline
		\end{tabular}
		\caption{
    Average effective sample size for the regression coefficients ($\mathcal{E}_{\bm \beta}$, compilation time in seconds ($t_c$) and iterations per second ($N_{it} / t_s$) for the linear model \eqref{eq:mix_lik} under different priors. From top to bottom: conjugate prior \eqref{eq:lm_conj}, conjugate prior with binary covariates, weakly informative prior \eqref{eq:lm_wi}, non informative prior  \eqref{eq:lm_ni} and lasso prior \eqref{eq:lm_lasso}. For each setting, values of $\mathcal{E}$ and $N_{it}/t_s$ associated to the software we recommend are highlighted in bold. Values of $(N_{it}, N_b, N_s)$ vary with the prior: (\num{11000}, \num{1000}  \num{5000}) for \eqref{eq:lm_conj},   (\num{15000}, \num{5000}, \num{5,000}) for \eqref{eq:lm_ni} and \eqref{eq:lm_wi} and   (\num{20000}, \num{10000},  \num{5000}) for \eqref{eq:lm_lasso}.
    For the \eqref{eq:lm_lasso}, superscripts attached to the value of $p$ 
    indicate the number of regression coefficients  $\beta_{j,true}$ set equal to zero.}
		\label{tab:lm_res}
	\end{table}

Table~\ref{tab:lm_res} here and Table~\ref{tab:lm_gof} in Appendix~\ref{sec:appendix_gof} show the indexes related to quality, efficiency and goodness of fit, as introduced in Section~\ref{sec:indexes}, for the linear model under different priors. 

When considering the conjugate prior distribution \eqref{eq:lm_conj}, both for continuous and binary covariates (see the first two blocks in Table~\ref{tab:lm_res}) JAGS and Stan have the highest \emph{quality} of the chain $\mathcal{E}$, represented by the highest $\mathcal{E}$. JAGS should be preferred as it is faster than Stan; see Table~\ref{tab:lm_res}. In this case, on the contrary, the default sampler chosen by NIMBLE produces highly autocorrelated chains resulting in low $\mathcal{E}$. NIMBLE is by far the fastest software when considering the number of iteration per second and we believe that with an appropriate fine-tuning of the MCMC strategy, NIMBLE can become competitive with JAGS and Stan. However, this kind of fine-tuning requires a deep understanding of the MCMC algorithms involved and might not be easy for practitioners.

Under  the weakly informative prior \eqref{eq:lm_wi}, JAGS and Stan  approximately get the same \emph{quality} of the chains, with JAGS being faster when the number of covariates is small and Stan being faster when it is large. In this case, the \emph{quality} of NIMBLE's chains is much better than under the conjugate prior, but still significantly smaller than for JAGS or Stan.

Under  the non-informative prior \eqref{eq:lm_ni}, performances are similar to those obtained for the conjugate prior, but when we consider a large dataset ($n$=1000, $p$=16), Stan becomes faster than JAGS, since it is able to generate 133 samples per second, while JAGS only 99. When $n<p$ ($n$=30, $p$=50), the sampling turns out to be difficult for all the software programs, and only Stan is able to generate chains with acceptable \textit{ess} ($\mathcal E$ around 78\%), even if it is much slower than JAGS and NIMBLE.

\begin{figure}
		\centering
		\begin{subfigure}{0.32\textwidth}
			\centering
			\includegraphics[width=\textwidth]{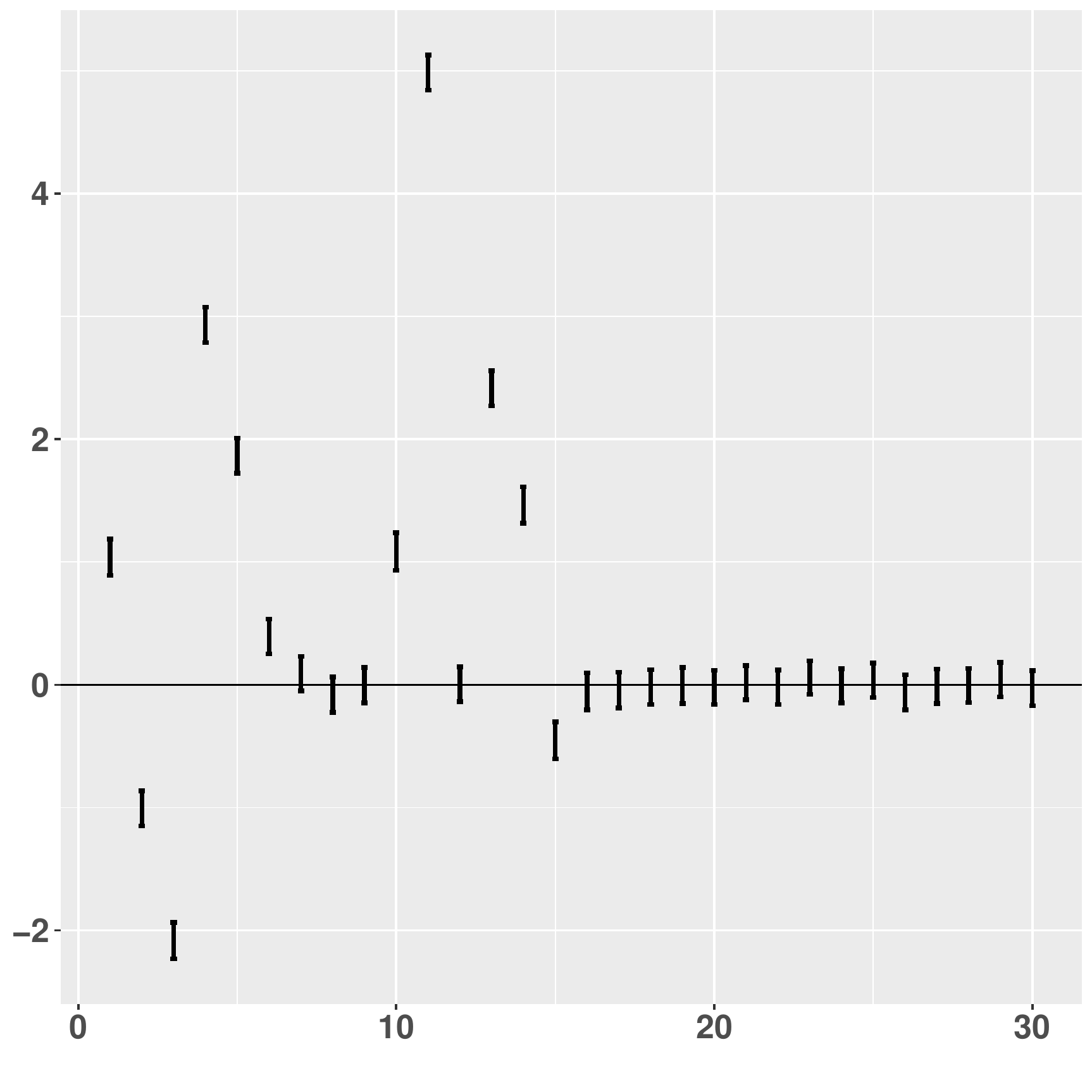}
			\caption{JAGS}
			
		\end{subfigure}
		\hfill
		\begin{subfigure}{0.32\textwidth}
			\centering
			\includegraphics[width=\textwidth]{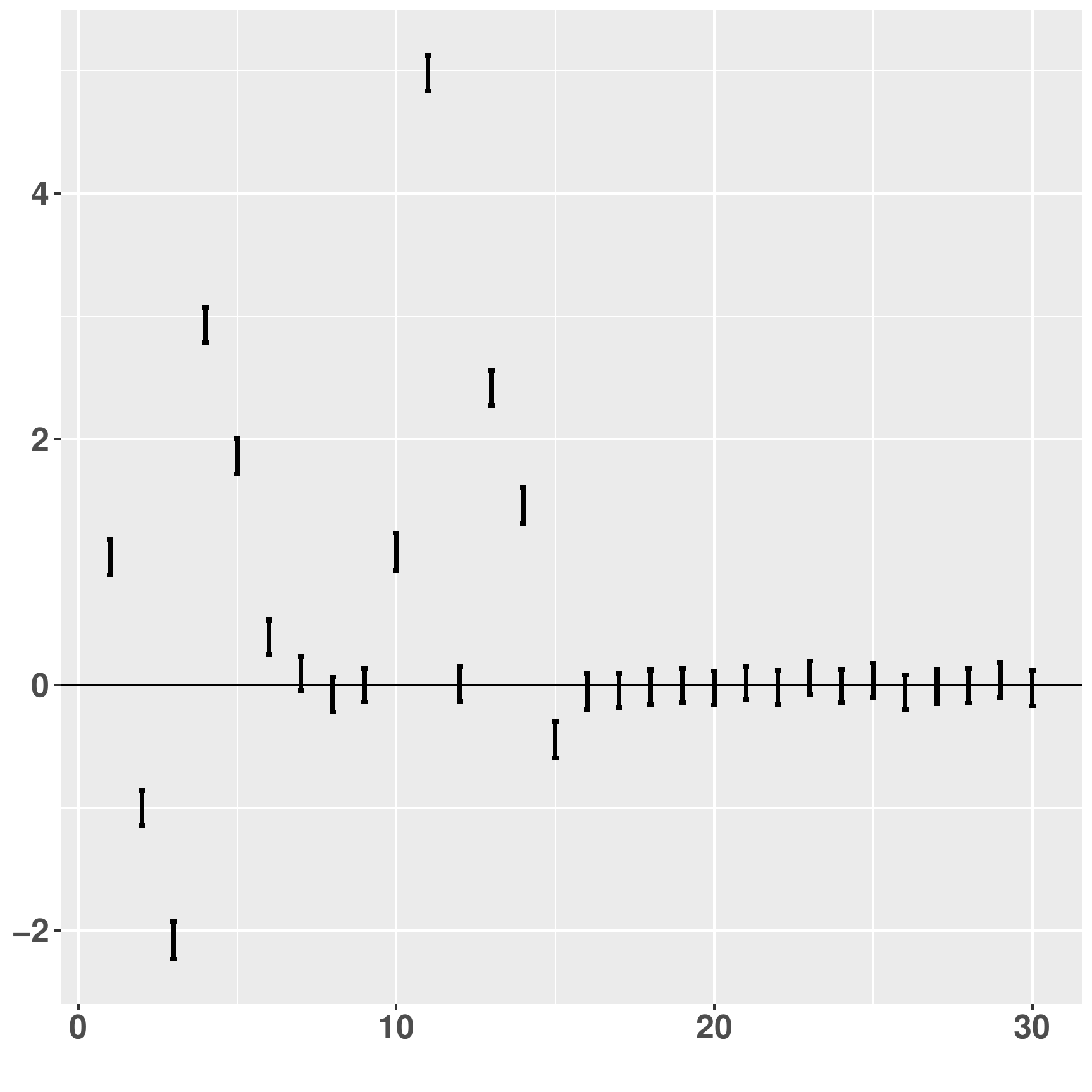}
			\caption{Stan}
			
		\end{subfigure}
		\hfill
		\begin{subfigure}{0.32\textwidth}
			\centering
			\includegraphics[width=\textwidth]{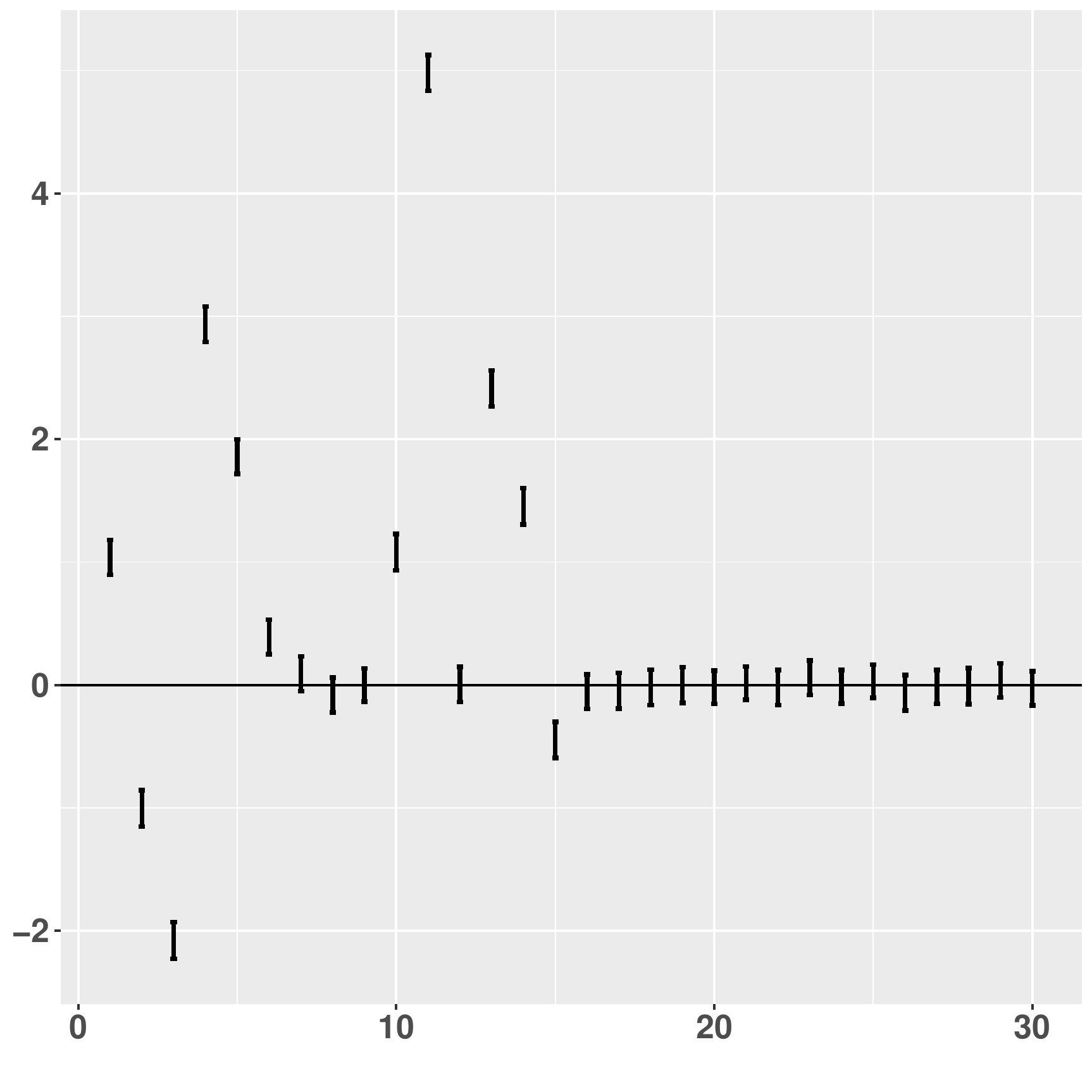}
			\caption{NIMBLE}
			
		\end{subfigure}
		\caption{95 \% credible intervals of the marginal posterior of each regression parameter for the linear model under lasso prior \eqref{eq:lm_lik}-\eqref{eq:lm_lasso} with $n$=1000 and $p$=30. The last 15 intervals on the right correspond to $\beta_{j,true}$=0.}
		\label{lasso_ic}
	\end{figure}

When considering  the lasso prior \eqref{eq:lm_lasso}, JAGS  turns out to be much slower than Stan and NIMBLE, and for this reason  we do not consider it in the comparison when $p$=100  and, in some cases, when $p=30$ (see Table~\ref{tab:lm_res}). Stan  is always able to generate almost uncorrelated chains, while the \textit{ess} of NIMBLE is around 35-40\% of the number of sampling iterations.  
Since, in addition, Stan is faster than NIMBLE, our final recommendation is for Stan in case of the lasso prior.
If we make variable selection through hard shrinkage, i.e., discarding covariates associated to coefficients $\beta_j$'s for which $95\%$ credible intervals of the marginal posterior contain the value 0,  
Figure~\ref{lasso_ic} shows that
this is consistent with the true data generating process when $n=1000$ and $p=30$ for every software platforms.
Although not reported here, similar conclusions hold for the other values of $p$ and $n$ considered in our examples.

The goodness-of-fit comparison in Table~\ref{tab:lm_gof} in Appendix~\ref{sec:appendix_gof} shows that both LPML and the errors are almost the same across the three software programs, meaning that the chains always converge to the same target distribution.

\paragraph{Logistic model}
	
For the logistic regression model \eqref{eq:lr_lik}, Table~\ref{tab:lr_res} shows that Stan is the  \virgolette{winner} since it guarantees both the highest \emph{quality} and the fastest sampling time, except when $p=4$, $n=100$ under prior \eqref{eq:lr_normal} when NIMBLE is faster but produces more autocorrelated chains.
Note that for all the platforms the $ess$ is much lower than  in case of the linear model. This can be imputed to a more complicated geometry of the posterior.

Under  the lasso prior \eqref{eq:lr_lasso}, our evidence is in line with what we have reported for the linear model. JAGS has not been considered when $p=100$ due to excessive runtimes. Moreover, the variable selection procedure is always consistent with the data generating process, the only exception we  have observed is in the case  of $p$=100 when we set 98 coefficients of $\bm \beta$ equal to zero: in this case, one interval from Stan and two intervals from NIMBLE did not contain the true zero value.

As far as goodness of fit is concerned,   JAGS offers a poorer fit to the data than Stan or NIMBLE. In particular the marginal expected value of the $\beta_j$'s obtained under JAGS are very different from the values used to generate the data when $p=16$. Stan and NIMBLE performance, from this point of view, are comparable (see Table~\ref{tab:lr_gof} in Appendix~\ref{sec:appendix_gof}).

\begin{table}[htbp]
		\centering 
		\caption*{\Large{Logistic Model}}
			\begin{tabular}{c | cc | cc | ccc |ccc}\hline
				& & &\multicolumn{2}{c|}{\textbf{JAGS}} &\multicolumn{3}{c|}{\textbf{STAN}} &\multicolumn{3}{c}{\textbf{NIMBLE}} \\
				& n & p & $\mathcal{E}_{\bm \beta}$ & $N_{it} / t_s$ & $t_c$ & $\mathcal{E}_{\bm \beta}$ & $N_{it} / t_s$  & $t_c$ & $\mathcal{E}_{\bm \beta}$ &$N_{it} / t_s$	\\ \hline
 				\parbox[t]{5mm}{\multirow{4}{*}{\rotatebox[origin=c]{90}{\eqref{eq:lr_normal}}}}
				& 100&4&30\%&1,000&130&\textbf{70\%} & \textbf{2,857} &51&15\%&4,000\\
				& 1000&4&24\%&72&130&\textbf{72\%} & \textbf{417} & 53&12\%&377\\
				& 100&16&12\%&217&130&\textbf{79\%} & \textbf{1,429} &40&13\%&1,115\\
				& 1000*&16*&7\%&14&130&\textbf{50\%} & \textbf{171} &65&7\%&100\\ \hline
				
				\parbox[t]{5mm}{\multirow{5}{*}{\rotatebox[origin=c]{90}{\eqref{eq:lr_lasso}}}}
				& 100&$16^{(0)}$&27\%&286&140&\textbf{63\%} & \textbf{2,000} &64&13\%&1,429\\
				& 1000&$16^{(0)}$&38\%&15&140& \textbf{77\%} & \textbf{345} &78&19\%&130\\
				& 1000&$100^{(2)}$&-&-&140&\textbf{28\%} & \textbf{68} &240&6\%&17\\
				& 1000*&$100^{*(50)}$&-&-&140& \textbf{63\%} & \textbf{57} &215&14\%&17\\
				& 1000&$100^{(98)}$&-&-&140&\textbf{91\%} & \textbf{41} &212&37\%&16\\\hline
		\end{tabular}
		\caption{Average effective sample size for the regression coefficients ($\mathcal{E}_{\bm \beta}$), compilation time in seconds ($t_c$) and iterations per second ($N_{it} / t_s$) for the logistic model \eqref{eq:lr_lik} under the different priors. From top to bottom: normal prior \eqref{eq:lr_normal} and lasso prior \eqref{eq:lr_lasso}. For each setting,  values of $\mathcal{E}$ and $N_{it}/t_s$ associated to the software we recommend 
		are highlighted in bold. 
    $(N_{it}, N_b, N_s)$ is equal to (\num{20000}, \num{10000},  \num{5000}) for \eqref{eq:lr_normal} and to $(15,000, \ 10,000, \ 2,500)$ for \eqref{eq:lr_lasso}.
For the \eqref{eq:lr_lasso}, superscripts attached to the value of $p$  
    indicate the number of regression coefficients  $\beta_{j,true}$ set equal to zero.}
		\label{tab:lr_res}
\end{table}
	
\paragraph{Mixture model}

\begin{table}[htbp]
		\centering 
		\caption*{\Large{Mixture Model}}
			\begin{tabular}{ cc | cc | ccc |ccc}\hline
				& &\multicolumn{2}{c|}{\textbf{JAGS}} &\multicolumn{3}{c|}{\textbf{STAN}} &\multicolumn{3}{c}{\textbf{NIMBLE}} \\
				 n & H & $\mathcal{E}_{v}$ & $N_{it} / t_s$ & $t_c$ & $\mathcal{E}_{v}$ & $N_{it} / t_s$  & $t_c$ & $\mathcal{E}_{v}$ &$N_{it} / t_s$	\\ \hline
				100&2&100\%&909&140&55\%&800&64&\textbf{100\%}&\textbf{2,500} \\
				1000*&2*&100\%&78&140&70\%&104&91&\textbf{100\%}&\textbf{250} \\
				100&4&43\%&500&140&78\%&83&70&\textbf{75\%} & \textbf{1,538}\\
				1000*&4*&81\%&44&140&60\%&6&93&\textbf{100\%} & \textbf{141} \\ \hline
		\end{tabular} \caption{Average effective sample size for the common coefficient ($\mathcal{E}_{v}$), compilation time in seconds ($t_c$) and iterations per second ($N_{it} / t_s$) for the mixture model \eqref{eq:mix_lik} under prior \eqref{eq:mix_prior}. For each setting,  values of $\mathcal{E}$ and $N_{it}/t_s$ associated to the software we recommend 
		are highlighted in bold.
		Here $(N_{it}, N_b, N_s)=$ (\num{20000}, \num{10000},  \num{5000}).}
		\label{tab:mix_res}
	\end{table}

The  label-switching issues mentioned in Section~\ref{sec:bay_mod} causes a severe non-identifiability  issue of the parameters in mixture models. Therefore, it would make no sense to compute the $ess$ of $\bm p, \bm \mu$ and $\bm \sigma^2$. Furthermore, in our simulations, none of the software is able to \virgolette{jump} between the modes of the posterior. Hence, we monitor only the chain of the common parameter $v^2$.

\begin{figure}
		\centering
		\begin{subfigure}{0.32\textwidth}
			\centering
			\includegraphics[width=\textwidth]{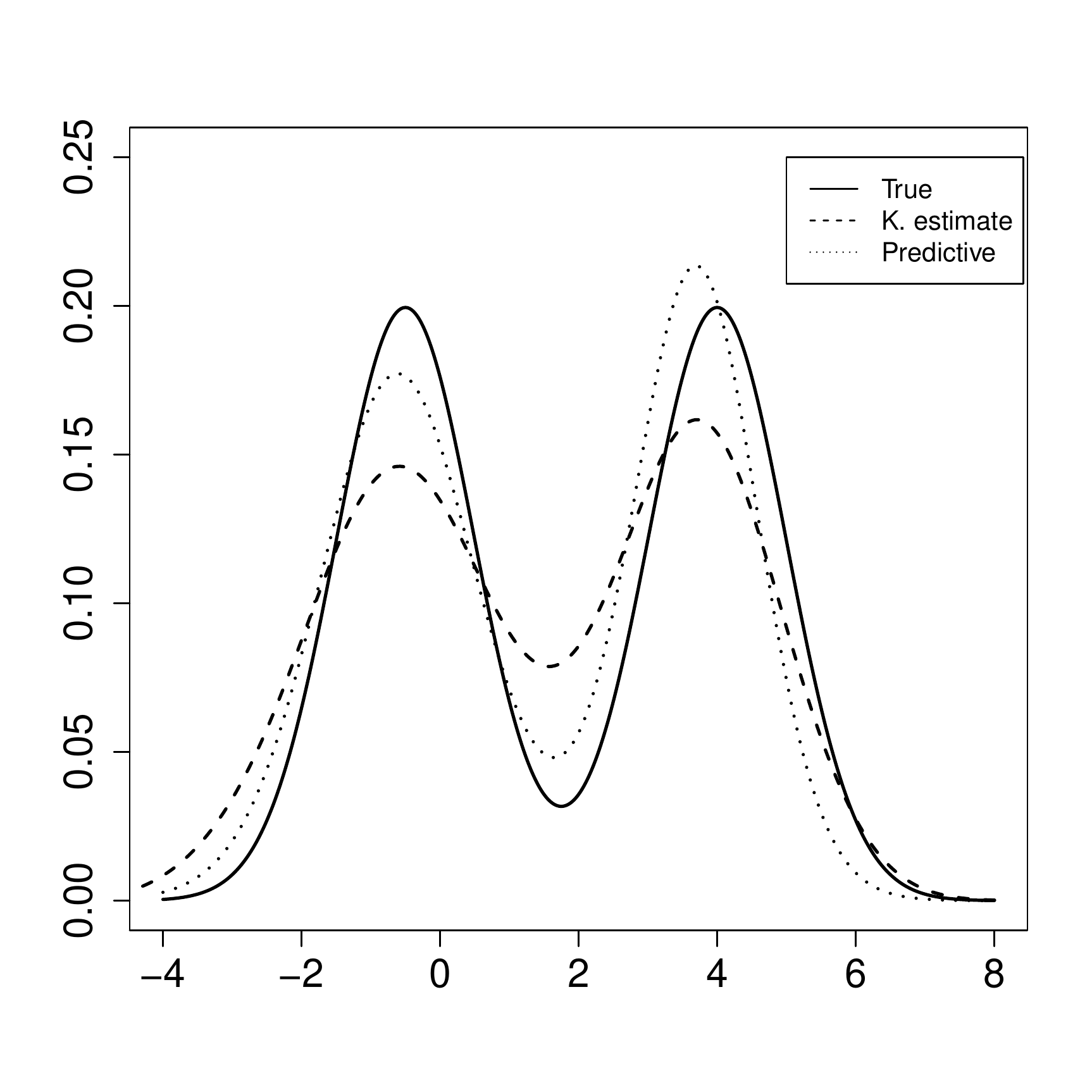}
			\caption{JAGS}
			
		\end{subfigure}
		\hfill
		\begin{subfigure}{0.32\textwidth}
			\centering
			\includegraphics[width=\textwidth]{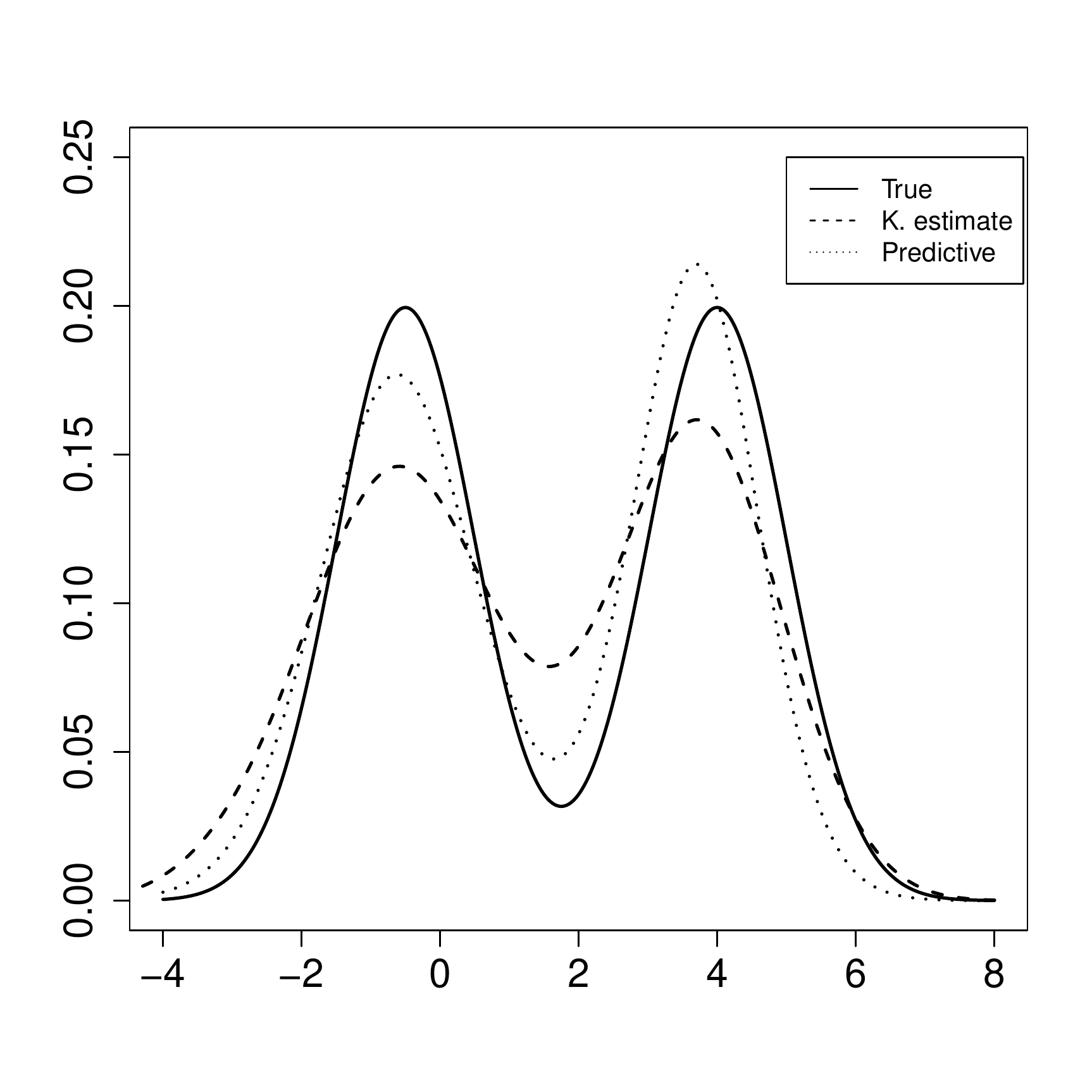}
			\caption{Stan}
			
		\end{subfigure}
		\hfill
		\begin{subfigure}{0.32\textwidth}
			\centering
			\includegraphics[width=\textwidth]{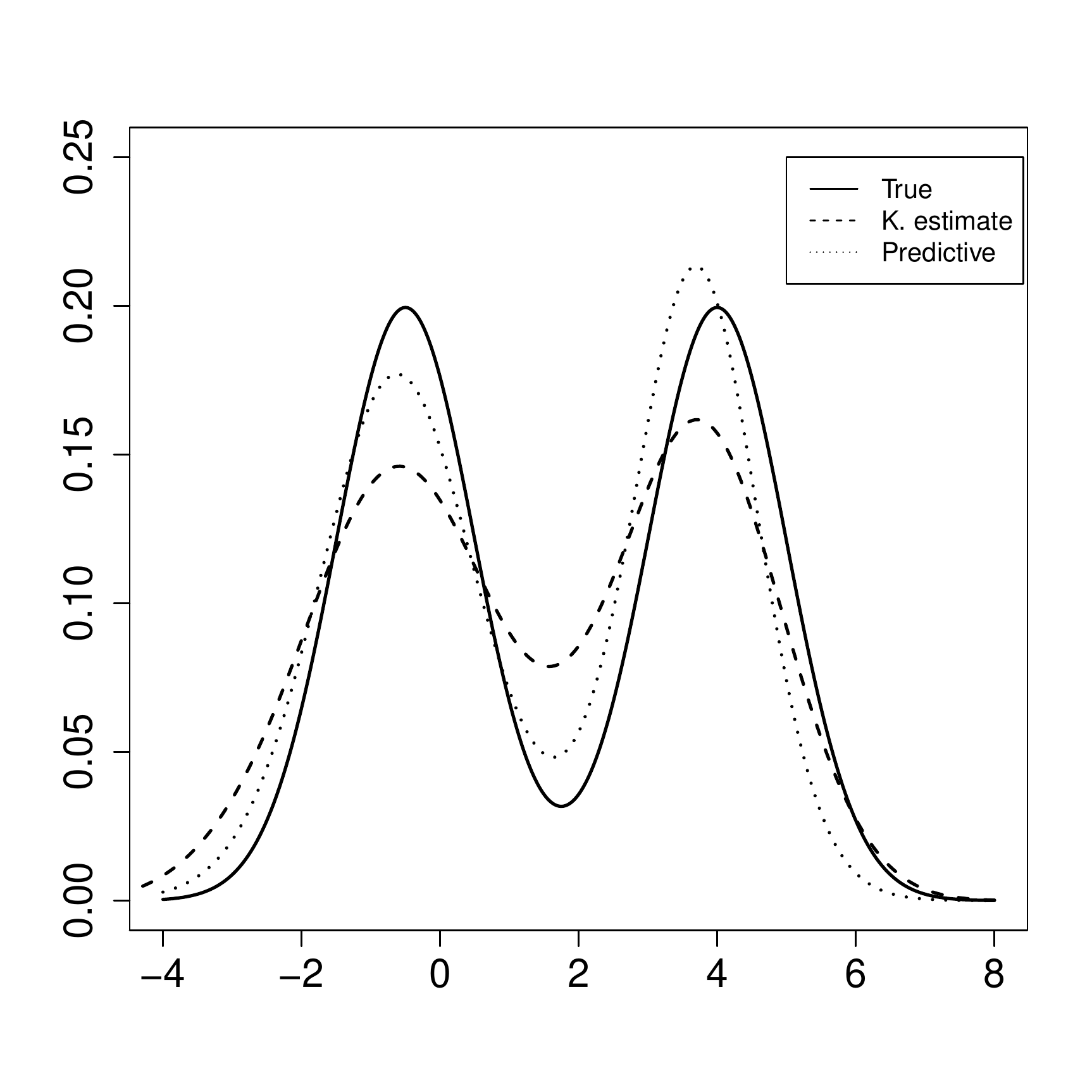}
			\caption{NIMBLE}
			
		\end{subfigure}
		\caption{Mixture models \eqref{eq:mix_lik}-\eqref{eq:mix_prior}, $H=2$, $n$=100: density generating the data (black line), kernel density estimate (dashed line), posterior predictive distributions (dotted line).}
		\label{mixture_predittiva_h2}
	\end{figure}

\begin{figure}
		\centering
		\begin{subfigure}{0.32\textwidth}
			\centering
			\includegraphics[width=\textwidth]{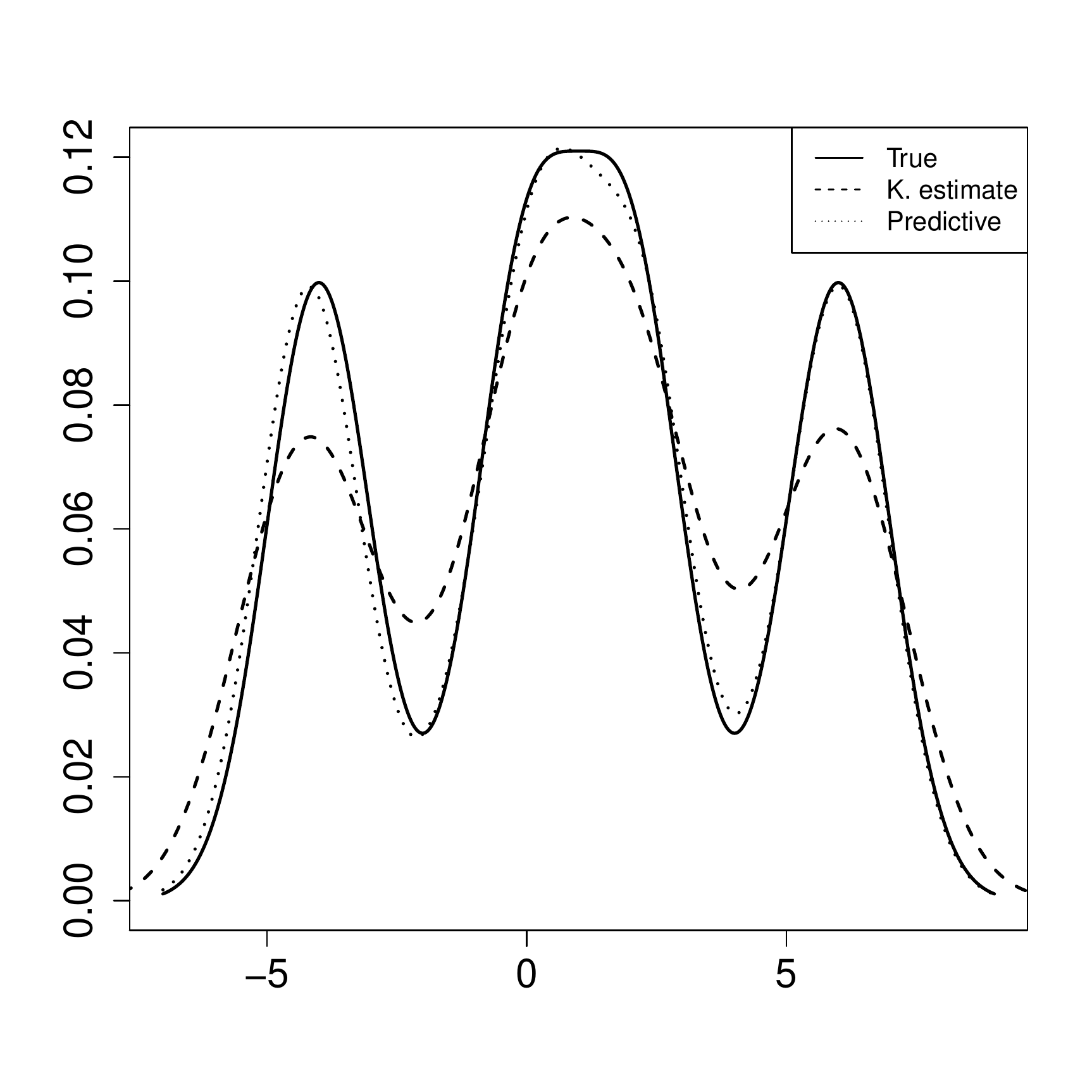}
			\caption{JAGS}
			
		\end{subfigure}
		\hfill
		\begin{subfigure}{0.32\textwidth}
			\centering
			\includegraphics[width=\textwidth]{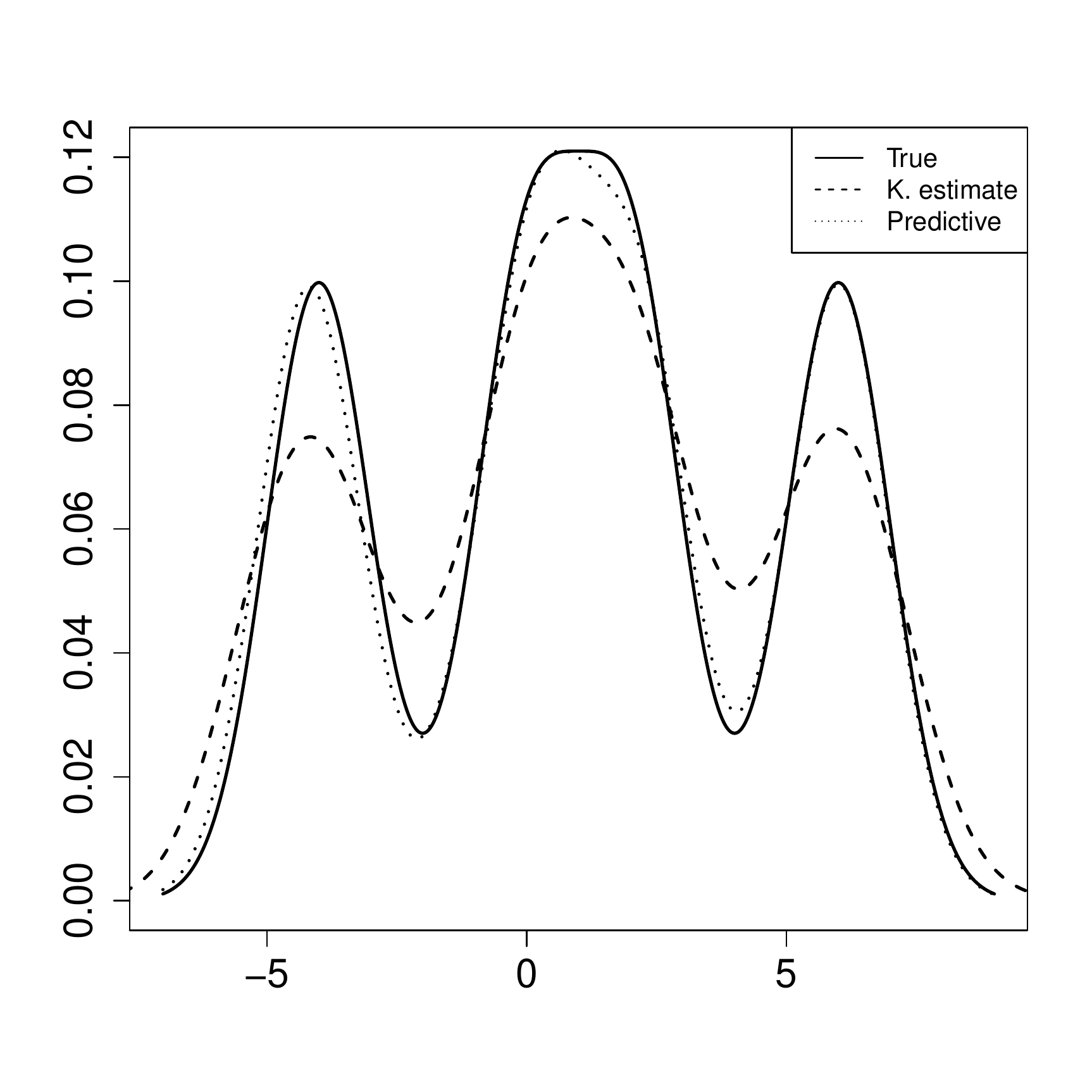}
			\caption{Stan}
			
		\end{subfigure}
		\hfill
		\begin{subfigure}{0.32\textwidth}
			\centering
			\includegraphics[width=\textwidth]{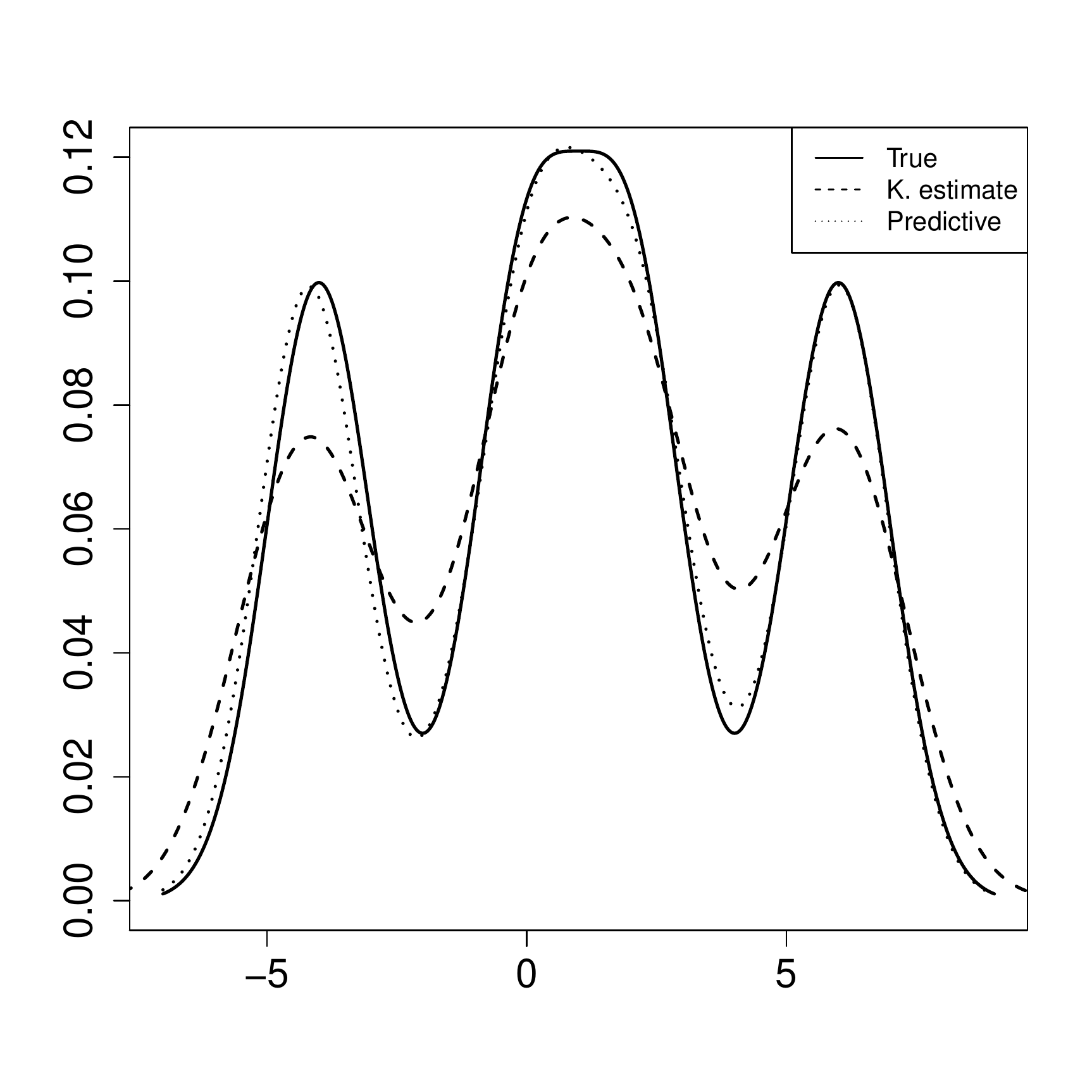}
			\caption{NIMBLE}
			
		\end{subfigure}
		\caption{Mixture models \eqref{eq:mix_lik}-\eqref{eq:mix_prior}, $H=4$, $n$=1000: density generating the data (black line), kernel density estimate (dashed line), posterior predictive distributions (dotted line).}
		\label{mixture_predittiva_h4}
	\end{figure}

Table~\ref{tab:mix_res} shows that 
NIMBLE is the fastest software and it also provides very high values of \textit{ess}, while JAGS and Stan present smaller values especially when $H$=4.
Figures~\ref{mixture_predittiva_h2} and \ref{mixture_predittiva_h4} ($H=2$ and $H=4$, respectively) show a comparison between the density used to generate the data, the kernel density estimate obtained from the data  (using the \texttt{density} function in  R) and the posterior predictive distribution obtained from the output of the software. 
The posterior predictive distributions are always accurate, for all the PPLs, and provide a better estimate of the data generating density than the classical kernel density estimate.
This is confirmed by Table~\ref{tab:mix_gof} in Appendix~\ref{sec:appendix_gof}, which shows WAIC indexes and Kullback-Leibler divergence between the predictive and the data generating densities. Looking at these scores, there is not much difference across the software, except when $n$=100 and $H$=4, where Stan present worst values.

\paragraph{Accelerated failure time model}

As for the accelerated failure time model \eqref{eq:aft_2} with the prior \eqref{eq:aft_nh}, Table~\ref{tab:aft_res} shows that JAGS and NIMBLE generate highly autocorrelated chains and only MCMC chains by Stan give high \textit{ess}. 
We observe similar performances when considering the non-informative prior \eqref{eq:aft_ni}. 
Analyzing the errors between the posterior means of the parameters and their true values (see Table~\ref{tab:aft_gof} in Appendix~\ref{sec:appendix_gof}), it is clear that, while the errors of JAGS and Stan are very similar, those of NIMBLE are larger. 
Moreover, Stan is the fastest software. For example, in the case of the non-hierarchical prior, considering 20\% of censored data, when $n$=1000 and $p$=16, Stan is able to generate 175 samples per second, while JAGS and NIMBLE only 10 and 65 respectively. For all these reasons, we recommend the use of Stan for AFT models with censored data.

\begin{table}[t]
		\centering 
		\caption*{\Large{AFT Model}}
			\begin{tabular}{cc | cc | cc | ccc |ccc}\hline
				& & & &\multicolumn{2}{c|}{\textbf{JAGS}} &\multicolumn{3}{c|}{\textbf{STAN}} &\multicolumn{3}{c}{\textbf{NIMBLE}} \\
				& & n & p & $\mathcal{E}_{\bm \beta}$ & $N_{it} / t_s$ & $t_c$ & $\mathcal{E}_{\bm \beta}$ & $N_{it} / t_s$  & $t_c$ & $\mathcal{E}_{\bm \beta}$ &$N_{it} / t_s$	\\ \hline
 				\parbox[t]{5mm}{\multirow{4}{*}{\rotatebox[origin=c]{90}{\eqref{eq:aft_nh}}}}
 				\parbox[t]{5mm}{\multirow{4}{*}{\rotatebox[origin=c]{90}{20\% C.}}}
				&& 100&4&67\%&476&115&\textbf{100\%}&\textbf{1167}&57&27\%&2,000\\
				&& 1000&4&69\%&41&115&\textbf{100\%}& \textbf{250}&72&6\%&200\\
				&& 100&16&46\%&135&115&\textbf{88\%}& \textbf{1111} &57&15\%&625\\
				&& 1000&16&71\%&10&115&\textbf{77\%}&\textbf{175}&85&5\%&65 \\\hline
				\parbox[t]{5mm}{\multirow{4}{*}{\rotatebox[origin=c]{90}{\eqref{eq:aft_nh}}}}
				\parbox[t]{5mm}{\multirow{4}{*}{\rotatebox[origin=c]{90}{50\% C.}}}
				&& 100&4&34\%&435&115&\textbf{97\%}&\textbf{1667}&54&4\%&2,000\\
				&& 1000&4&48\%&34&115&\textbf{92\%}&\textbf{303}&79&2\%&200\\
				&& 100&16&23\%&133&115&\textbf{92\%}&\textbf{1111}&54&8\%&588\\
				&& 1000*&16*&-&-&115&\textbf{70\%}&\textbf{179}&90&2\%&60 \\\hline
				\parbox[t]{5mm}{\multirow{4}{*}{\rotatebox[origin=c]{90}{\eqref{eq:aft_nh} }}}
				\parbox[t]{5mm}{\multirow{4}{*}{\rotatebox[origin=c]{90}{80\% C.}}}
				&& 100&4&10\%&455&115&\textbf{74\%}&\textbf{1667}&51&2\%&1,818\\
				&& 1000&4&16\%&41&115&\textbf{97\%}&\textbf{400}&76&1\%&169\\
				&& 100&16&7\%&169&115&\textbf{68\%}&\textbf{1111}&59&5\%&526\\\
				&& 1000&16&-&-&115&\textbf{79\%}&\textbf{333}&92&1\%&69 \\\hline
				\parbox[t]{5mm}{\multirow{4}{*}{\rotatebox[origin=c]{90}{\eqref{eq:aft_ni}}}}
				\parbox[t]{5mm}{\multirow{4}{*}{\rotatebox[origin=c]{90}{50\% C.}}}
				&& 100&4&35\%&476&130&\textbf{95\%}&\textbf{1,667}&58&3\%&2,000\\
				&& 1000&4&47\%&39&130&\textbf{100\%}&\textbf{227}&78&3\%&196\\
				&& 100&16&23\%&130&130&\textbf{89\%}&\textbf{1,000}&62&8\%&556\\
				&& 1000&16&43\%&8&130&\textbf{82\%}&\textbf{152}&95&2\%&52 \\\hline
		\end{tabular}
		\caption{Average effective sample size for the regression coefficients ($\mathcal{E}_{\bm \beta}$), compilation time in seconds ($t_c$) and iterations per second ($N_{it} / t_s$) for the accelerated failure time model \eqref{eq:aft_2} under the different priors and different percentage of censored data. From top to bottom: non hierarchical prior \eqref{eq:aft_nh} considering 20\%, 50\% and 80\% of censored data and non-informative prior \eqref{eq:aft_ni} considering 50\% of censored data. For each setting, 
values of $\mathcal{E}$ and $N_{it}/t_s$ associated to the software we recommend
		are highlighted in bold.
		$(N_it, N_b, N_s)$ amount to  (\num{10000}, \num{5000}, \num{2500}).}
		\label{tab:aft_res}
	\end{table}
	
\subsection{Parallel chains}
\label{sec:parallel_chains}

MCMC is an inherently sequential procedure, 
so that parallelization of the code usually does not produce significant speed-ups in terms of sampling time.
However, multiple chains can be run in parallel and independently and then their output can be combined. 
For models marked with * in Tables~\ref{tab:lm_res}-\ref{tab:aft_res} we also  tested the performance of the software platforms when running four independent Markov chains in parallel. 
 Below we report a qualitative summary of our findings. 
	
As far as the \emph{quality} is concerned, except for some small differences due to Monte Carlo variability, we observed that the $\mathcal{E}$ statistics are unchanged with respect to the single-chain setup of Section~\ref{sec:comparison}. 
Regarding the sampling times, JAGS takes slightly more time with respect to the single chains simulations, but overall the parallelization is very efficient. Stan and NIMBLE in general require the same time as the single chains simulations and, in some cases, they are even faster.
Hence, running four chains in parallel usually results in a speed-up of a factor four when considering  the statistics \emph{ess} / $t_s$.

As one might expect, the amount of memory used by the software increases linearly with the number of cores (i.e., parallel chains).
JAGS is the more memory-parsimonious software: even when all 8 cores were used, the amount of RAM required was less than 1 GB.
Regarding Stan and NIMBLE, we should separately take compilation phase and  sampling phase into account.
In Stan, the model is compiled once (possibly using more than one core depending on the machine configuration) and the same compiled model is used independently by all the parallel chains. This requires approximately 2 GB of memory.
In NIMBLE instead,  each core must separately re-compile the model: as NIMBLE manual says \virgolette{\textit{This ensures that all models and algorithms are independent objects that do not interfere with each other.}} \citep{nimble_manual}.
As a result, the compilation phase in NIMBLE is very demanding in terms of memory, with each C++ compiler requiring around 350 MB, while further memory is required by RStudio. 
Overall, when eight parallel chains are generated, the compilation phase in NIMBLE requires up to 7 GB. 
For all the models tested, the sampling phase requires less memory than the compilation phase, that is around 150 MB per core using Stan (with additional 600-800 MB of memory required by Rstudio) and around 350 MB per core using NIMBLE (plus 800-900 MB used by Rstudio).
Overall, these figures  are easily handled by modern laptops even when using six or more chains.

As an example, in Table~\ref{memory_lm} we report the amount of memory used by the software for a linear model under conjugate prior when the number of parallel chains is 1, 4 and 8.
	
	\begin{table}
		\centering
		
		\begin{tabular}{c| cc | ccc}\hline
			\textbf{Chains}\hspace{2pt} &\multicolumn{2}{c|}{\textbf{Compilation phase} } &\multicolumn{3}{c}{\textbf{Sampling phase}}
			\\
			& \textbf{STAN} & \textbf{NIMBLE} & \textbf{JAGS}& \textbf{STAN} & \textbf{NIMBLE}\\ \hline
			1& 1,950&1,600 &550&500&1,200\\
			4& 1,950&3,400&650&1,100&2,000  \\
			8&1,950&6,200&800&1,600&3,400 \\\hline
			
		\end{tabular}
		\caption{Amount of memory (in MB) required by JAGS, Stan and NIMBLE for computing MCMC of all the parameters of a conjugate linear model with 1,4,8 chains.}
		\label{memory_lm}
	\end{table}
	
Our suggestion is to use parallelization, as it greatly saves   computational time, even if it is more demanding in terms of memory. As Kruschke (\citeyear{Kruschke}) suggests, if the computer has $K$ cores, it is better to run at most $K-1$ parallel chains and to reserve the remaining core for other tasks.

\subsection{Main takeaways and general guidelines}

In this section we would like to offer a more general point of view on the output of the PPLs, also taking into account the different MCMC algorithms each platform uses, and to provide general guidelines  about the platform that is  \textit{best} suited according to the model at stake.

Taking into account all of our simulations, Stan appears the go-to solution if one wants to learn only one between JAGS, Stan and NIMBLE programs. 
In fact, the NUTS HMC sampler employed by Stan provides always high-quality chain (measured by the index $\mathcal{E}$). 
Since HMC relies on the gradient of the log posterior density, considering non-informative priors such as the uniform distribution, for which the gradient is computationally cheap to compute, provides a speed-up for the sampling time required by Stan.
However, Sections~\ref{sec:software_comparison} and \ref{sec:parallel_chains} report cases when JAGS or NIMBLE are preferable. The greatest limitation of Stan, as underlined by many authors,  is the unfeasibility of discrete parameters, as the computation of the gradient in this case is not allowed.

In general, conjugate or semi-conjugate models (i.e., when the full-conditionals are in closed form) are very well fitted by JAGS, which is usually faster than Stan and it is able to generate chains with high effective sample size.
  In fact, since in conjugate or semi-conjugate models full conditional distributions are obtained in closed form, the Gibbs sampler used by JAGS  samples from them easily and quickly. 
As a result, JAGS is faster than Stan, which is penalized by the computation of the gradient that requires some time.  For further software comparison in case of semi-conjugate models, see also \cite{falcothesis}.

NIMBLE performs MCMC simulations via the Random Walk Metropolis Hastings or using the Gibbs sampler, when the model is recognized to be semi-conjugate.
In our examples, chains produced using Random Walk were highly correlated. Further, conjugacy or semi-conjugacy is not always recognized by NIMBLE; for instance, we  have checked that posterior MCMC simulation of the conjugate linear model \eqref{eq:lm_lik}-\eqref{eq:lm_conj} was performed through a Random Walk.
Instead, NIMBLE has proved to be the best software to fit mixture models, where the Gibbs sampler was used,  in this case being the fastest software and the one with highest quality of the chains, especially in the case of four components mixtures.
We also remark that, for mixture models, Stan requires the extra effort of writing down the likelihood in the marginal form \eqref{eq:mix_lik}, unlike NIMBLE and JAGS.

NIMBLE has also proved to be faster (in terms of runtime) than Stan and JAGS in several simulations. An expert user could benefit from this speed and get less autocorrelated chains by controlling the MCMC sampling scheme of NIMBLE. Indeed,  NIMBLE is the only software that allows the control of the MCMC algorithm, though this opportunity requires knowledge of MCMC algorithms that might not be accessible to practitioners.
	
We have separately reported compilation times and sampling times by Stan and NIMBLE, since compilation occurs before sampling for both platforms. Compilation phase of Stan requires around 2-3 minutes without significant differences between the models we analysed. Compilation phase of NIMBLE takes usually the same amount of time, i.e. 2-3 minutes, though this time reduces for small datasets.

\section{Discussion}
\label{sec:discussion}

Motivated by the personal need to use  PPLs able to quickly derive MCMCs for computing the posterior, in this work we have compared the performance of three software platforms, i.e.,   JAGS, NIMBLE and Stan. 
These  PPLs are able to automatically generate samples from the posterior distribution of interest using MCMC algorithms, starting from the specification of a Bayesian model, i.e. the likelihood and the prior.
Our extensive simulation studies evaluated the quality of the MCMC chains produced, the efficiency of the software and the goodness of fit of the output. We tested several Bayesian models using synthetic datasets, varying the sample size and the dimension of the parameter space.
We also considered the efficiency of the parallelization made by the three platforms.

Of course,  quantitative comparisons of probabilistic programming languages as those considered here are
vastly available on the web in the form of blog posts or tutorials. While these resources are undoubtedly useful, they usually present analyses only on particular case studies or class of models which results into going through dozens of blog posts. 
We do not deny the usefulness of these webpages or blogs, but we believe that
general results on this comparison should be rearranged, synthesised and made available to the scientific community in the form of a scientific article, as we have done with this manuscript.

From  our detailed analysis of the  MCMCs and the study of the MCMC algorithms of each platforms, our conclusion is that  Stan is the default go-to software. However, if the model contains latent discrete parameters that cannot be analytically marginalized out, contrary to the case of mixture models that allow for marginalization, then 
Stan is ruled out by the impossibility of computing the gradient.
NIMBLE is usually very fast, but the use of the Random Walk  Metropolis Hastings algorithm might entail highly autocorrelated chains. However, expert users can benefit from the modularity of NIMBLE and change the default sampling strategy, although we have not investigated more deeply here.
JAGS is very efficient when the models are semi-conjugate, that is when the full-conditionals are available in closed form. Moreover, with the exception of mixture models, our evidence shows that the default sampler from JAGS is typically more efficient than the default sampler from NIMBLE.

In recent years, the use of graphical processing units (GPUs) has contributed to the popularity of deep learning methods to solve several challenging real-world problems using massive datasets. GPUs can be used to parallelize computation, providing speed-ups of several order of magnitudes with respect to CPU processing.
Although MCMC is inherently sequential, new PPLs have been developed to work on GPUs, building on existing deep learning library \citep[see, e.g.,][]{tran2016edward, bingham2019pyro, dillon2017tensorflow}. These new PPLs aim  either at parallelizing the code within the single chain, hence promising to help scaling MCMC to large data, or at sampling hundreds of parallel chains. To take full advantage of GPUs, alternatives to the NUTS algoritm are also being developed \citep{hoffman2021adaptive}.
However, all these new PPLs are based on Python, which is the default programming language for deep learning, while R is still the most popular language among applied statisticians. For this reason, in our comparison, we have only considered PPLs with an R interface.

\appendix
	
	\section{Technical details on MCMC algorithms}
	\label{sec:details}
	
Several inferential objectives can be expressed as the expected value of a function $f: \Theta \rightarrow \mathbb{R}$ with respect to the posterior density $\pi(\bm \theta \mid \bm y)$, i.e.
\begin{equation}\label{eq:post_int}
\mathbb{E}[f(\bm \theta)\mid \bm y] = \int_\Theta f(\bm \theta) \pi(\bm \theta \mid \bm y) \mathrm{d} \bm \theta.
\end{equation}
For instance, using $f(\bm \theta) = v_i(\bm \theta) = (\theta_i - \bar{\theta}_i)^2$ in \eqref{eq:post_int} amounts to computing the posterior 
marginal variance of a parameter $\theta_i$ ($\bar{\theta}_i$ being the posterior mean).
Given MCMC samples $\bm \theta^{(1)}, \bm \theta^{(2)}, \ldots, \bm \theta^{(N_s)}$ with limit/invariant density $\pi(\bm \theta \mid \bm y)$, Monte Carlo integration approximates \eqref{eq:post_int}  by
\begin{equation}\label{eq:mc_int}
    \mathbb{E}[f(\bm \theta)\mid \bm y] \approx Q_{N_s}(f) = \frac{1}{N_s} \sum_{j=1}^{N_s} f(\bm \theta^{(j)}).
\end{equation}

If $\bm \theta^{(1)}, \bm \theta^{(2)}, \ldots, \bm \theta^{(N_s)}$ were iid samples from the posterior distribution (i.e. the posterior can be sampled exactly and not using MCMC), then the law of large number would guarantee that $Q_{N_s}(f)$ converges to $\mathbb{E}[f(\bm \theta)\mid \bm y]$ as $N_s \rightarrow +\infty$. The central limit theorem characterizes the speed of convergence, since
${\displaystyle    \frac{Q_{N_s}(f) - \mathbb{E}[f(\bm \theta)]}{\sqrt{\text{Var}(f(\bm \theta)) / N_s}}}$ converges in distribution to the standard Gaussian r.v. as $N_s$ goes to $+\infty$. In this case,
the Monte Carlo estimator is unbiased for any value of $N_s$, i.e. $\mathbb{E}[Q_{N_s}(f)] = \mathbb{E}[f(\bm \theta)\mid \bm y]$.

However, in practice, it is never the case that the $\bm \theta^{(j)}$'s from the MCMC are independent, since they are realizations from a Markov chain. 
Fortunately, suitable versions of the law of large number and the central limit theorem hold true also in this case, so that $Q_{N_s}(f)$ still converges to the \virgolette{true} value $ \mathbb{E}[f(\bm \theta)\mid \bm y]$  almost surely. However, in case of non-iid MCMC samples,
the estimator $Q_{N_s}(f)$ is biased for every choice of $N_s$ (i.e., the bias disappears only when $N_s = +\infty$) and the speed of convergence becomes smaller than  as in the independent case. 
In particular, the bias term comes from the initialization, unless the initial value $\bm \theta^{(1)}$ is sampled exactly from the posterior. The bias is usually ignored, since using a reasonable number of  \virgolette{burn-in} iterations and having a large enough $N_s$ makes it negligible.  The burn-in becomes relevant if one wants to run several short chains in parallel. 
Interested readers may read \cite{jacob2020unbiased} for a recent development on this issue.

Given the Markov dynamics, the law of $\bm \theta^{(i+1)}$ depends on $\bm \theta^{(i)}$ and this dependency  is propagated through all the subsequent samples. The dependency between the MCMC samples will likely yield (positive) autocorrelations between $\{f(\bm \theta^{(j)})\}$, i.e.
\[
  \tau^f_{i, j} := \text{Cov}(f(\bm \theta^{(i)}, f(\bm \theta^{(j)}) > 0
\]
for most $i,j$. 
If the chain has reached stationarity, then $\tau^f_{i, j}$ depends only on the \emph{lag} between $i, j$, that is $\tau^f_{i, j} = \tau^f_{\ell}$ where $\ell = j - i$.
The coefficient $\rho^f_ \ell = \tau^f_{\ell} / \text{Var}(f(\bm \theta)\mid \bm y)$ is called $\ell$-lag \emph{autocorrelation}. It can be shown that the autocorrelation is symmetric in $\ell$ ($\tau^f_{\ell} = \tau^f_{-\ell}$) and $\tau^f_0 = 1$. The central limit theorem for Markov chains states that
\[
    \frac{Q_{N_s}(f) - \mathbb{E}[f(\bm \theta)\mid \bm y]}{\sqrt{
    \frac{\text{Var}(f(\bm \theta)\mid \bm y)}{  {N_s}/{ \sum_{\ell=-\infty}^{\infty} \rho^f_\ell} }}} \rightarrow \mathcal{N}(0, 1) \textrm{ as } N_s\rightarrow +\infty,
\]
i.e. the left hand side converges in distribution to the standard Gaussian r.v..
The difference between the \virgolette{ideal} rate of convergence of $Q_{N_s}$ (under iid samples) and the actual rate, obtained with MCMC samples, is measured in terms of \emph{effective sample size}, that is 
$N_s  / \sum_{\ell=-\infty}^{\infty} \rho^f_\ell $. Informally, the effective sample size quantifies the amount of information contained in a sample of size $N_s$ from an MCMC in terms of the number of independent samples that contain the same information.

\begin{figure}
    \centering
    \includegraphics[width=\linewidth]{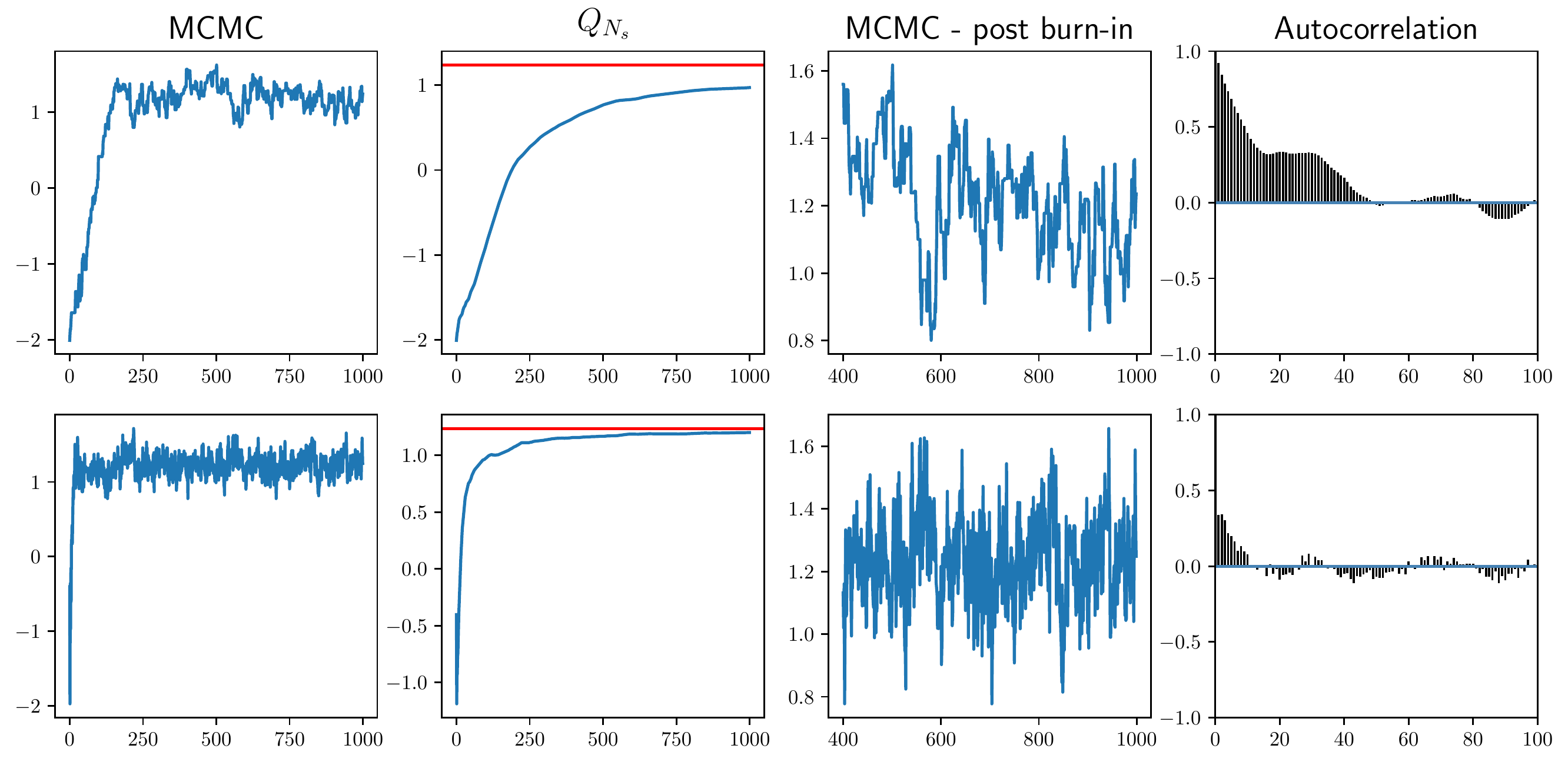}
    \caption{Graphical comparison of two MCMC. Top row MCMC was obtained by Random Walk Metropolis Hastings, the  bottom row MCMC  by HMC. Columns from left to right: full MCMC path, performance of $Q_{N_s}(f)$ with $f(x) = x$ (i.e. posterior mean) as a function of $N_s$, zoom-in on the last 600 iterations and relative autocorrelations computed on the last 600 iterations. }
    \label{fig:mcmc_ex}
\end{figure}

Consequently, we aim at building MCMC algorithms with two features. First, the MCMC should reach stationarity quickly, so that, after discarding the burn-in iterations, the estimator $Q_{N_s}(f)$ shows little bias. Secondly, the MCMC should have low autocorrelations,  to guarantee that the convergence speed (equivalently, the variance of the estimator $Q_{N_s}(f)$) is comparable to that obtained in the case of simple Monte Carlo integration.
Figure~\ref{fig:mcmc_ex} shows two MCMC chains on a toy problem with $f(x) = x$. The top row MCMC reaches stationarity after roughly 250 samples and exhibits high autocorrelation, while the  bottom row MCMC is stationary almost immediately and has significantly lower autocorrelation.
Their difference is highlighted comparing the two 
estimators $Q_{N_s}$ (without burn-in) in the second column in Figure~\ref{fig:mcmc_ex}: when $N_s = 1000$ the top row estimator is still far from the correct value (horizontal line) while the  bottom row estimator approaches the true value after only 300 iterations.

\section{Glossary}
\label{sec:appendix_distr}
	We report the probability densities used in the paper, specifying the notation adopted. When we specify the argument of the density function, this means to specify the support of the distribution.
However, we have not specified the range of parameter values in Table~\ref{tab:distributions}.
The normalizing constant  $B(\bm \alpha)$ of the Dirichlet distribution in Table~\ref{tab:distributions} is  the beta integral 
$$B(\bm{\alpha})=\frac{\Gamma(\alpha_1)\cdots\Gamma(\alpha_K)}{\Gamma(\alpha_1+\cdots+\alpha_K)} . $$ 

	\begin{table}[h]
		\begin{tabular}{ll}
			\hline
			Gaussian distribution & \multirow{2}*{$f(x|\mu, \sigma^2)=\frac{1}{\sqrt{2\pi\sigma^2}}\exp{\Big({-\frac{(x-\mu)^2}{2\sigma}}\Big)}$} \\
			
			$ X|\mu,\sigma^2  \sim \mathcal{N}(\mu,\sigma^2)$ \vspace{4pt}\\
			\hline
			$p$-variate Gaussian distribution & \multirow{2}*{$f( \bm{x}|\bm{\mu},\Sigma)=\frac{\exp{\big(-\frac{1}{2}(\bm{x}-\bm{\mu})^T \Sigma^{-1}(\bm{x}-\bm{\mu})\big)}}{\sqrt{(2\pi)^n det(\Sigma)}}$} \\
			
			$\bm{X}|\bm{\mu},\Sigma \sim \mathcal{N}_p(\bm{\mu},\Sigma)$ \vspace{4pt}\\
			\hline
			Uniform distribution & \multirow{2}*{$f(x|a,b)=\frac{1}{b-a}, \hspace{10pt} x \in [a,b]$} \\
			
			$ X|a,b \sim \mathcal{U}(a,b)$ \vspace{4pt}\\
			\hline
			Bernoulli distribution & \multirow{2}*{$f(x|p)=\begin{cases}
					1-p\ & \hbox{if} \ x=0\\
					p\ & \hbox{if} \ x=1
				\end{cases}$} \\
			
			$X|p \sim Be(p)$ \vspace{10pt}\\
			\hline
			Exponential distribution & \multirow{2}*{$ f(x|\lambda)=\lambda e^{-\lambda x}, \hspace{10pt} x>0$} \\
			
			$X|\lambda \sim \mathcal{E}(\lambda)$ \vspace{4pt}\\
			\hline
			
			Double-Exponential distribution & \multirow{2}*{$  f(x|\mu,b)=\frac{1}{2b}\exp{\Big({-\frac{|x-\mu|}{b}}\Big)}$} \\
			
			$X|\mu,b \sim \mathcal{DE}(\mu, b)$ \vspace{4pt}\\
			\hline
			Caucy distribution & \multirow{2}*{$  f(x|\mu,b)=\frac{1}{\pi}\frac{b}{(x-\mu)^2+b^2}$} \\
			
			$ X|\mu,b \sim \mathcal{C}(\mu,b)$ \vspace{4pt}\\
			\hline
			Half-Caucy distribution & \multirow{2}*{$  f(x|\mu,b)=\frac{2}{\pi}\frac{b}{(x-\mu)^2+b^2}, \hspace{10pt} x>\mu$} \\
			
			$ X|\mu,b \sim \mathcal{HC}(\mu,b)$ \vspace{4pt}\\
			\hline
			Inverse-Gamma distribution & \multirow{2}*{$f(x|\alpha,\beta)=\frac{\beta^{\alpha}}{\Gamma(\alpha)}\frac{e^{-\beta/x}}{x^{\alpha+1}}, \hspace{10pt} x>0$} \\
			
			$ X|\alpha,\beta \sim \mathcal{IG}(\alpha,\beta)$ \vspace{4pt}\\
			\hline
			Weibull distribution & \multirow{2}*{$f(x|\alpha,\lambda)=\alpha \lambda x^{\alpha -1}e^{-\lambda x^{\alpha}},\hspace{10pt} x>0$} \\
			
			$ X|\alpha,\lambda \sim \text{Wei}(\alpha,\lambda)$ \vspace{4pt}\\
			
			\hline
			
			Dirichlet distribution & $f(\bm{x}|\bm{\alpha})=\frac{1}{B(\bm{\alpha})}\prod_{i=1}^{K-1}x_{i}^{\alpha_i -1}(1-\sum_{i=1}^{K-1}x_{i})^{(\alpha_K -1)} $, \\ 
			
			$X_1,\dots,X_{K-1}|\alpha_1,\dots,\alpha_K \sim \mathcal{D}(\bm{\alpha})$ &
			$  x_i \in (0,1), \hspace{5pt} \sum_{i=1}^{K-1}x_{i}\leq 1$  \vspace{10pt}\\
			\hline
			Categorical distribution & \multirow{2}*{$f(x|\bm{p})=\begin{cases}
					p_i\ & \hbox{if} \ x=i; \hspace{10pt} i=1,\dots,K\\
					0 \ & \hbox{otherwise}
				\end{cases}$} \\
			
			$X|\bm{p} \sim \text{cat}(K;\bm{p})$ \vspace{10pt}\\
			
			\hline
		\end{tabular}
		\caption{Probability distributions.}
		\label{tab:distributions}
	\end{table}

\section{Monitoring goodness of fit}
\label{sec:appendix_gof}

In this section, for each model, we monitor posterior predictive goodness-of-fit indexes. In particular, we have computed LPML, WAIC and the
Kullback--Leibler divergence between the true and estimated densities. We also have considered the \virgolette{error} between the posterior mean of the parameters and their true value.  Specifically, denoting as $\bm{\theta}^{(j)}$ the value of the parameters at iteration $j=1,\dots,N_s$ and as $f(\cdot,\bm{\theta}^{(j)})$ the likelihood evaluated at iteration $j$, we have computed
	\begin{equation*}
		\text{LPML} =  \sum_{i=1}^n \log\left(\frac{1}{\frac{1}{N_{s}}\sum_{j=1}^{N_{s}}\frac{1}{f(y_i|\bm{\theta}^{(j)})}}\right)
	\end{equation*}
	\begin{equation*}
		\text{WAIC}=  \sum_{i=1}^n \log\left(\frac{1}{N_s}\sum_{j=1}^{N_s}f(y_i|\bm{\theta}^{(j)})\right) - \sum_{i=1}^n Var(\log \ f(y_i|\bm{\theta})) 
	\end{equation*}
See \cite{BIDA}	and \cite{waic} for the definition of LPML  and WAIC, respectively.
	
 For mixture models, posterior predictive inference  has been compared by the Kullback--Leibler divergence between the true distribution $p(y)$ generating the data and its Bayesian estimate, that is the posterior predictive distribution $q(y)$:
	\begin{equation*}
		\text{KL}(p||q)=  \int_{\mathbb{R}} p(y)\log_2 \left(\frac{p(y)}{q(y)}\right) dy.
	\end{equation*}
Better predictive performances are detected by higher LPML and WAIC and lower Kullback--Leibler divergence. 
	
		The \virgolette{error} is computed as the sum of the squares of the differences between the posterior mean of the regression parameters $\beta_j$'s, denoted by $\hat{\beta_j}$, and the true value used to simulate the data, denoted by $\beta_{j,true}$:
	\begin{equation*}
		\text{Error}=  \sum_{j=1}^p (\hat{\beta_j}-\beta_{j,true})^2.
	\end{equation*}

 Tables~\ref{tab:lm_gof}-\ref{tab:aft_gof} show the goodness-of-fit indexes and the error estimates for the posterior analyses in Section~\ref{sec:comparison}.
	
	\begin{table}[h!]
		\centering 
		\caption*{\Large{Linear Model}}
		\begin{tabular}{c | cc | cc | cc |cc}\hline
			& & &\multicolumn{2}{c|}{\textbf{JAGS}} &\multicolumn{2}{c|}{\textbf{STAN}} &\multicolumn{2}{c}{\textbf{NIMBLE}} \\
			& n & p  &LPML&Error&LPML&Error&LPML&Error	\\ \hline
			\parbox[t]{5mm}{\multirow{6}{*}{\rotatebox[origin=c]{90}{\eqref{eq:lm_conj}}}}
			&100&4&-233.92&0.41&-234.08&0.41&-234.24&0.38\\
			&1000&4&-2266.44&0.0044&-2266.34&0.0044&-2266.45&0.0039\\
			&100&16&-238.24&1.057&-238.05&1.056&-238.76&1.076\\
			&1000&16&-2243.56&0.089&-2243.48&0.090&-2243.08&0.099\\
			&2000&30&-4445.78&0.0488&-4445.69&0.0487&-4466.44&0.0495\\
			&30&50&-101.69&486.41&-103.479&487.24&-101.68&563.11\\ \hline
			\parbox[t]{5mm}{\multirow{5}{*}{\rotatebox[origin=c]{90}{\eqref{eq:lm_conj} - Bin}}}
			&100&4&-228.61&3.876&-228.69&3.939&-228.82&3.852\\
			&1000&4&-2266.33&0.0045&-2266.34&0.0044&-2266.45&0.0039\\
			&100&16&-240.76&29.81&-240.88&29.85&-240.88&25.63\\
			&1000&16&-2243.42&0.0891&-2243.48&0.0895&-2243.08&0.0994\\
			&30&50&-107.77&358.03&-107.08&354.62&-101.12&473.85\\ \hline
			\parbox[t]{5mm}{\multirow{4}{*}{\rotatebox[origin=c]{90}{\eqref{eq:lm_wi}}}}
			&100&4&-235.22&0.12&-235.23&0.12&-235.33&0.12\\
			&1000&4&-2235.02&0.024&-2235.09&0.024&-2235.09&0.024\\
			&100&16&-231.23&0.509&-230.96&0.509&-230.72&0.492\\
			&1000&16&-2212.30&0.060&-2212.53&0.0560&-2212.31&0.0591\\ \hline
			\parbox[t]{5mm}{\multirow{4}{*}{\rotatebox[origin=c]{90}{\eqref{eq:lm_ni}}}}
			&100&4&-235.25&0.121&-235.27&0.112&-235.32&0.117\\
			&1000&4&-2235.11&0.023&-2235.04&0.020&-2235.05&0.024\\
			&100&16&-230.89&0.509&-230.99&0.518&-230.96&0.509\\
			&1000&16&-2212.32&0.059&-2212.28&0.059&-2212.31&0.060\\
			&30&50&-185.51&5049.4&-192.37&288.6&-187.20&11823.2
			\\\hline 
			\parbox[t]{5mm}{\multirow{8}{*}{\rotatebox[origin=c]{90}{\eqref{eq:lm_lasso}}}}
			&100&$16^{(1)}$&-225.46&1.05&-225.37&1.06&-225.29&1.08\\
			&1000&$16^{(1)}$&-2213.38&0.068&-2213.47&0.068&-2213.39&0.068\\
			&1000&$30^{(2)}$&-&-&-2275.18&0.088&-2275.34&0.089\\
			&1000&$30^{(3)}$&-2274.63&0.086&-2273.91&0.087&-2274.18&0.087\\
			&1000&$30^{(4)}$&-&-&-2270.13&0.0541&-2270.09&0.0543\\
			&1000&$100^{(2)}$&-&-&-2284.90&0.397&-2284.95&0.395\\
			&1000&$100^{(3)}$&-&-&-2282.07&0.375&-2282.00&0.373\\
			&1000&$100^{(4)}$&-&-&-2254.49&0.1829&-2254.41&0.1815
			\\\hline
		\end{tabular}
		\caption{LPML and error of the posterior mean estimate of regression parameters for the linear model \eqref{eq:lm_lik} under different priors. The numbers of iterations $(N_{it}, N_b, N_s)$ are as in Table~\ref{tab:lm_res}.}
		\label{tab:lm_gof}
	\end{table}

	\begin{table}[t]
		\centering 
		\caption*{\Large{Logistic Model}}
	\begin{tabular}{c | cc | cc | cc |cc}\hline
		& & &\multicolumn{2}{c|}{\textbf{JAGS}} &\multicolumn{2}{c|}{\textbf{STAN}} &\multicolumn{2}{c}{\textbf{NIMBLE}} \\
		& n & p  &LPML&Error&LPML&Error&LPML&Error	\\ \hline
			\parbox[t]{5mm}{\multirow{4}{*}{\rotatebox[origin=c]{90}{\eqref{eq:lr_normal}}}}
			&100&4&-29.68&0.491&-28.96&0.215&-28.75&0.212\\
			&1000&4&-209.98&0.25&-209.85&0.16&-209.77&0.13\\
			&100&16&-151.71&599.15&-47.04&17.37&-49.95&18.12\\
			&1000&16&-131.11&16.99&-118.07&1.64&-118.05&1.88 \\ \hline
			
			\parbox[t]{5mm}{\multirow{5}{*}{\rotatebox[origin=c]{90}{\eqref{eq:lr_lasso}}}}
			&100&$16^{(1)}$&-26.37&4.495&-25.48&5.505&24.68&5.852\\
			&1000&$16^{(1)}$&-180.49&0.458&-180.44&0.464&-180.07&0.475\\
			&1000&$100^{(2)}$&-&-&-132.78&11.81&-132.17&14.40\\
			&1000&$100^{(3)}$&-&-&-158.23&4.49&-159.64&4.33\\
			&1000&$100^{(4)}$&-&-&-390.13&0.386&-389.80&0.384\\ \hline
		\end{tabular}
		\caption{LPML and error of the posterior mean estimate of regression parameters for the logistic regression model \eqref{eq:lr_lik} under different priors. The numbers of iterations $(N_{it}, N_b, N_s)$ are as in Table~\ref{tab:lr_res}.}
		\label{tab:lr_gof}
	\end{table}

\begin{table}[t]
	\centering 
	\caption*{\Large{Mixture Model}}
	\begin{tabular}{c | cc | cc | cc |cc}\hline
	& & &\multicolumn{2}{c|}{\textbf{JAGS}} &\multicolumn{2}{c|}{\textbf{STAN}} &\multicolumn{2}{c}{\textbf{NIMBLE}} \\
	& n & H  & WAIC & KL & WAIC & KL & WAIC & KL	\\ \hline
		&100&2&-210.80&0.0642&-210.76&0.0639&-210.88&0.0653\\
		&1000&2&-2,093.17&0.0015&-2,093.23&0.0015&-2,093.28&0.0015\\
		&100&4&-260.84&0.0622&-260.88&0.0774&-260.91&0.0631\\
		&1000&4&-2,585.93&0.0046&-2,582.91&0.0046&-2,583.25&0.0048\\ \hline
	\end{tabular} \caption{WAIC index and Kullback–Leibler divergence between the true distribution generating the data and the posterior predictive distribution obtained from the MCMC for the mixture model \eqref{eq:mix_lik}. The numbers of iterations $(N_{it}, N_b, N_s)$ are as in Table~\ref{tab:mix_res}.}
	\label{tab:mix_gof}
\end{table}

\begin{table}[t]
	\centering 
	\caption*{\Large{AFT Model}}
	\begin{tabular}{cc | cc | c | c |c}\hline
		& & & &\textbf{JAGS} &\textbf{STAN} &\textbf{NIMBLE} \\
		& & n & p & Error & Error & Error	\\ \hline
		\parbox[t]{5mm}{\multirow{4}{*}{\rotatebox[origin=c]{90}{\eqref{eq:aft_nh}}}}
		\parbox[t]{5mm}{\multirow{4}{*}{\rotatebox[origin=c]{90}{20\% C.}}}
		&&100&4&0.041&0.041&0.031\\
		&&1000&4&0.009&0.009&0.129\\
		&&100&16&0.506&0.510&0.454\\
		&&1000&16&0.021&0.022&0.383\\\hline
		\parbox[t]{5mm}{\multirow{4}{*}{\rotatebox[origin=c]{90}{\eqref{eq:aft_nh}}}}
		\parbox[t]{5mm}{\multirow{4}{*}{\rotatebox[origin=c]{90}{50\% C.}}}
		&&100&4&0.107&0.109&0.102\\
		&&1000&4&0.013&0.013&0.539\\
		&&100&16&1.641&1.712&3.564\\
		&&1000&16&-&0.040&1.980\\\hline
		\parbox[t]{5mm}{\multirow{4}{*}{\rotatebox[origin=c]{90}{\eqref{eq:aft_nh} }}}
		\parbox[t]{5mm}{\multirow{4}{*}{\rotatebox[origin=c]{90}{80\% C.}}}
		&&100&4&0.752&0.576&0.324\\
		&&1000&4&0.010&0.010&5.593\\
		&&100&16&8.67&11.27&6.59\\
		&&1000&16&-&0.050&4.462\\\hline
		\parbox[t]{5mm}{\multirow{4}{*}{\rotatebox[origin=c]{90}{\eqref{eq:aft_ni}}}}
		\parbox[t]{5mm}{\multirow{4}{*}{\rotatebox[origin=c]{90}{50\% C.}}}
		&&100&4&0.101&0.100&0.100\\
		&&1000&4&0.013&0.013&0.546\\
		&&100&16&1.763&7.686&3.603\\
		&&1000&16&0.042&0.043&1.759\\\hline
	\end{tabular}
	\caption{Error of the posterior mean estimate of regression parameters for the accelerated failure time model \eqref{eq:aft_2} under different priors. The numbers of iterations $(N_{it}, N_b, N_s)$ are as in Table~\ref{tab:aft_res}.}
	\label{tab:aft_gof}
\end{table}

\section{Repeated simulations}
\label{sec:repeated_simulations}
	
To guarantee robustness of conclusions, some models were tested many times  (20, 30 or 50 times). 
Posterior MCMC simulations have been performed using different datasets generated as described in Section~\ref{subse:simulation_data}. 
	In particular, the linear model under the conjugate prior, the logistic regression model under the normal prior and the AFT model under non hierarchical prior have been repeatedly tested  considering the sample size $n$=1000 and the number of parameters $p$=8. For instance data used to test the LM with the conjugate prior have been simulated from the linear model \eqref{eq:lm_lik} with different fixed values of $\bm{\beta}$, $\sigma^2$ and $X$. 
Posterior MCMC simulations of the AFT models have been repeated varying also the percentage of censored data (20\%, 50\% and 80\%). 
	For the linear model under the lasso prior we have fixed $n$=800 and $p$=30, while half of true values of the regression parameters were set equal to zero. MMs were tested considering $n$=800 and $H$=2. 
	
\begin{figure}
		\centering
		\begin{subfigure}{1\textwidth}
			\centering
			\includegraphics[width=0.66\textwidth]{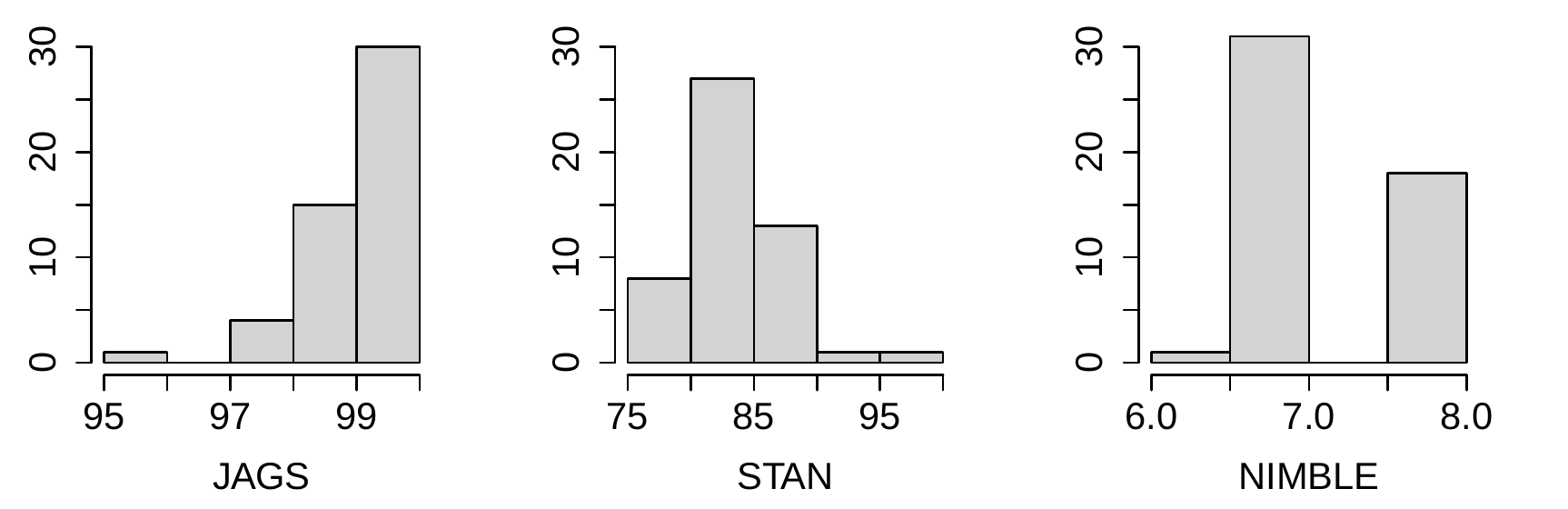}
			\caption{\textit{ess $\beta$}/$N_{s}$ expressed in \%.}
			
		\end{subfigure}
		\hfill
		\vspace{15pt}\\
		\begin{subfigure}{1\textwidth}
			\centering
			\includegraphics[width=0.66\textwidth]{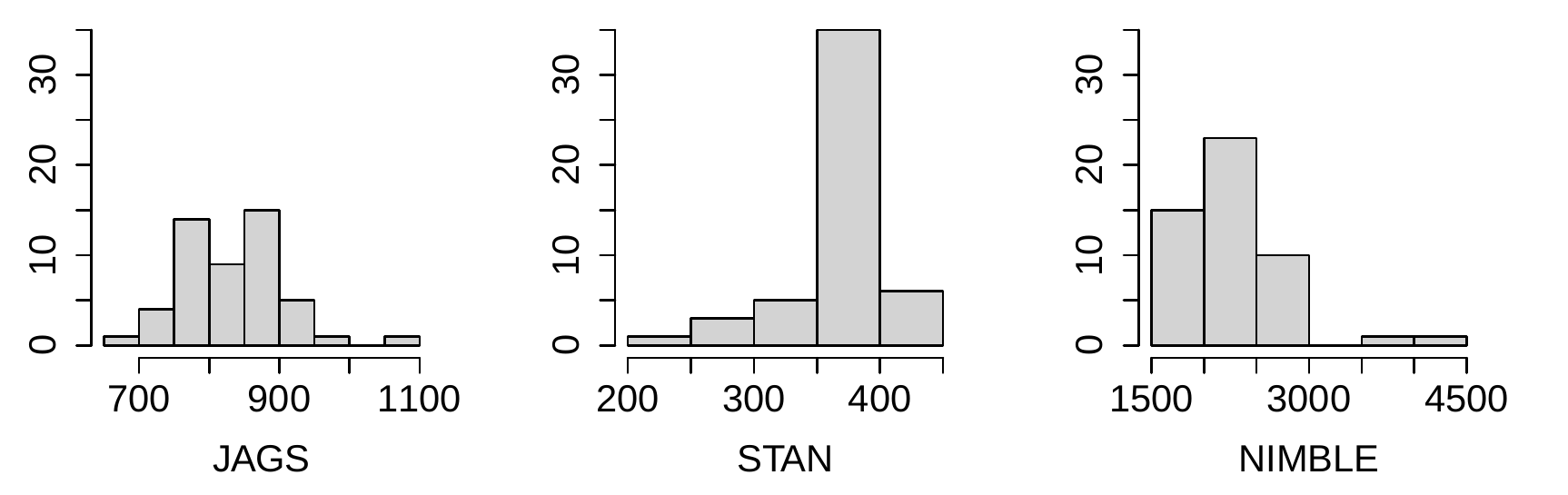}
			\caption{$N_{it}$/$t_{s}$}
			
		\end{subfigure}
		\caption{Histograms of \textit{ess$\beta$}/$N_{s}$ (a) and $N_{it}$/$t_{s}$ (b) over 50 simulated datasets ($n$=1000, $p$=8) for the linear model under prior \eqref{eq:lm_conj}. }
		\label{fig:lm_conj_artiglieria}
	\end{figure}

\begin{figure}
		\centering
		\begin{subfigure}{1\textwidth}
			\centering
			\includegraphics[width=0.66\textwidth]{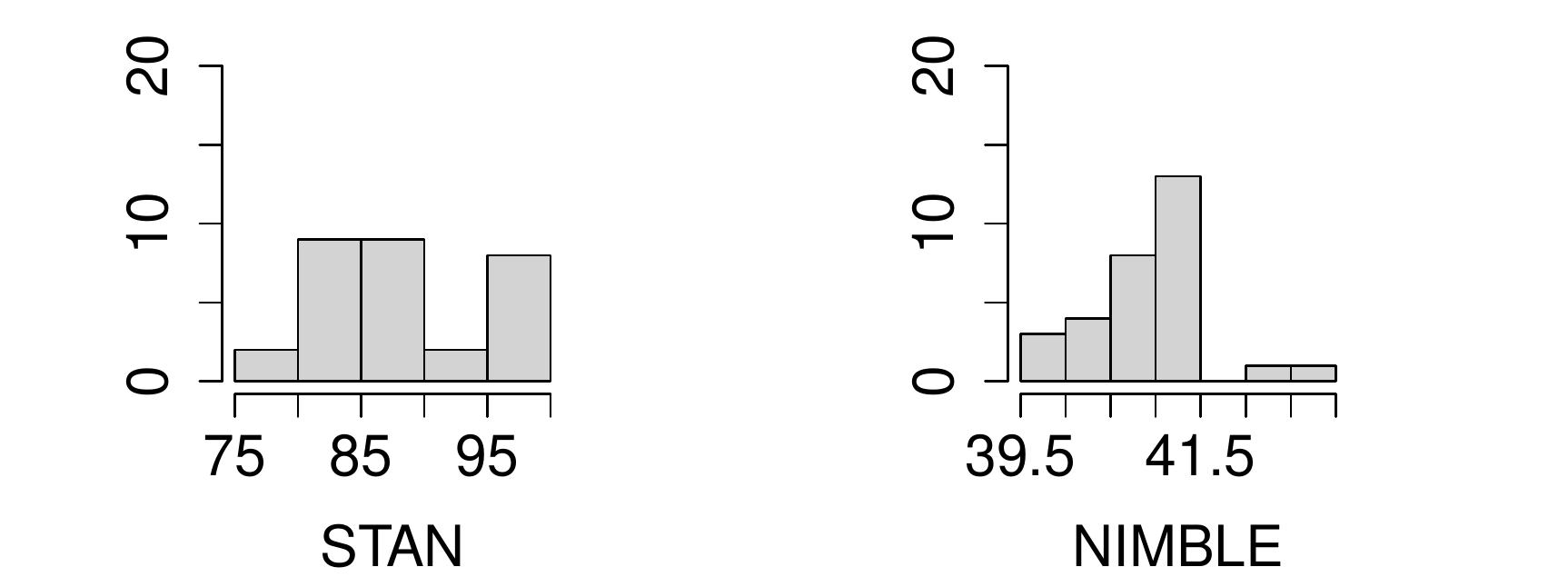}
			\caption{\textit{ess $\beta$}/$N_{s}$ expressed in \%.}
			
		\end{subfigure}
		\hfill
		\vspace{15pt}\\
		\begin{subfigure}{1\textwidth}
			\centering
			\includegraphics[width=0.66\textwidth]{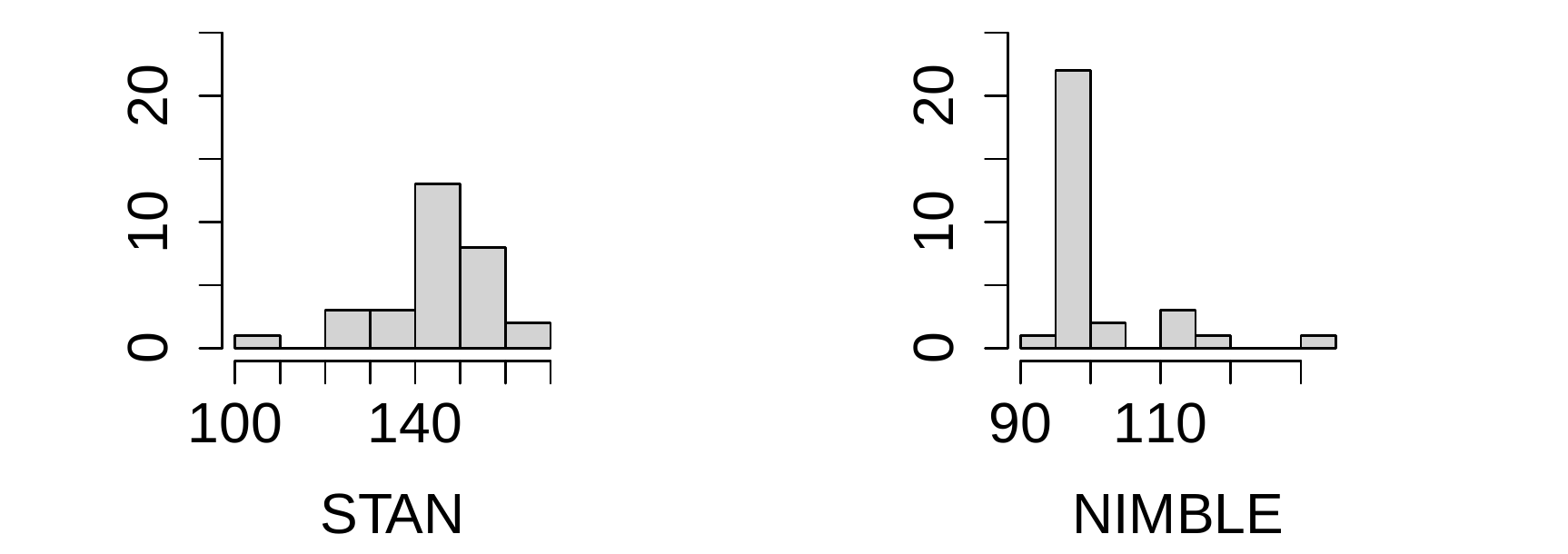}
			\caption{$N_{it}$/$t_{s}$}
			
		\end{subfigure}
		\caption{Histograms of \textit{ess$\beta$}/$N_{s}$ (a) and $N_{it}$/$t_{s}$ (b) over 30 simulated datasets ($n$=800, $p$=30) for the linear model under prior \eqref{eq:lm_lasso}.}		
		\label{fig:lm_lasso_artiglieria}
	\end{figure}

\begin{figure}
		\centering
		\begin{subfigure}{1\textwidth}
			\centering
			\includegraphics[width=0.66\textwidth]{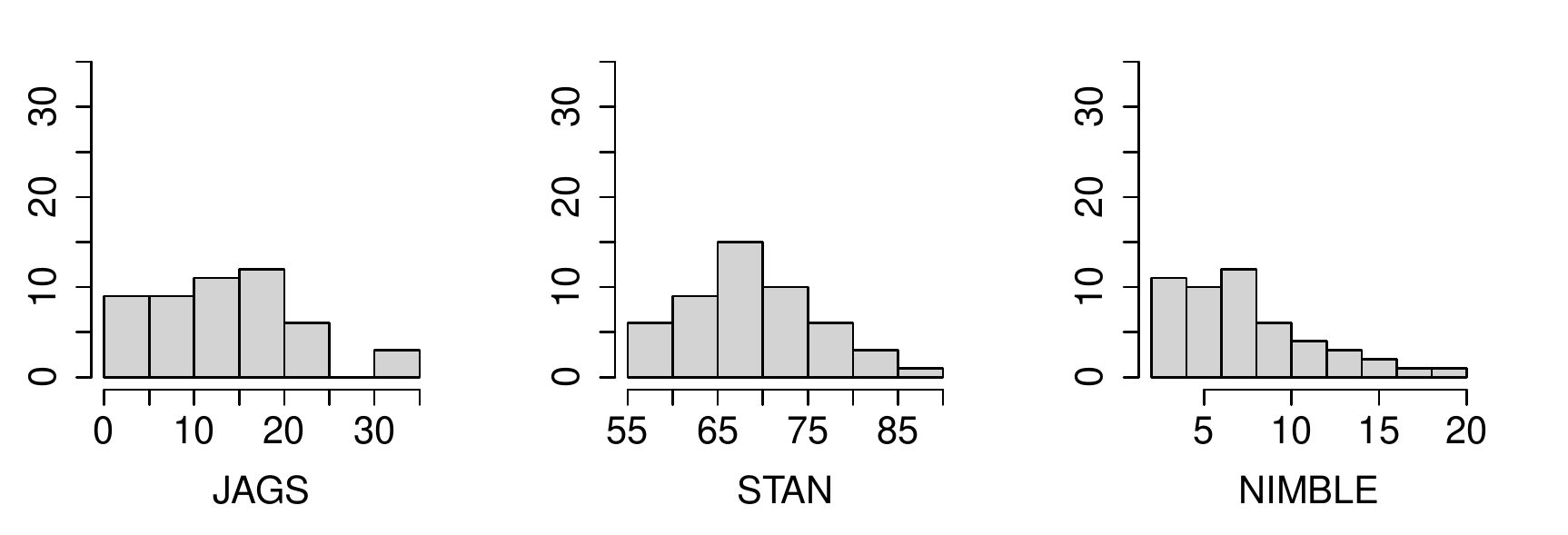}
			\caption{\textit{ess $\beta$}/$N_{s}$ expressed in \%.}
			
		\end{subfigure}
		\hfill
		\vspace{15pt}\\
		\begin{subfigure}{1\textwidth}
			\centering
			\includegraphics[width=0.66\textwidth]{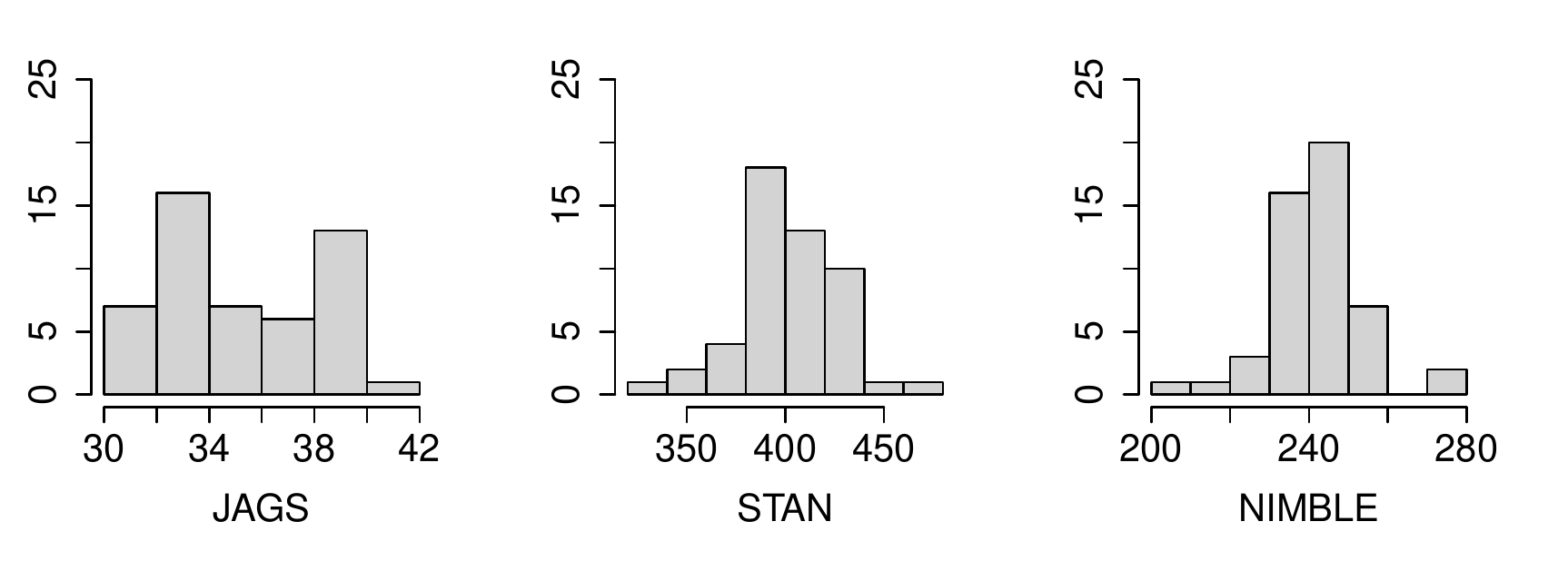}
			\caption{$N_{it}$/$t_{s}$}
			
		\end{subfigure}
		\caption{Histograms of \textit{ess$\beta$}/$N_{s}$ (a) and $N_{it}$/$t_{s}$ (b) over 50 simulated datasets ($n$=1000, $p$=8) for the logistic regression mode under prior \eqref{eq:lr_normal}.
}
		\label{fig:logreg_artiglieria}
	\end{figure}

\begin{figure}
		\centering
		\begin{subfigure}{1\textwidth}
			\centering
			\includegraphics[width=0.66\textwidth]{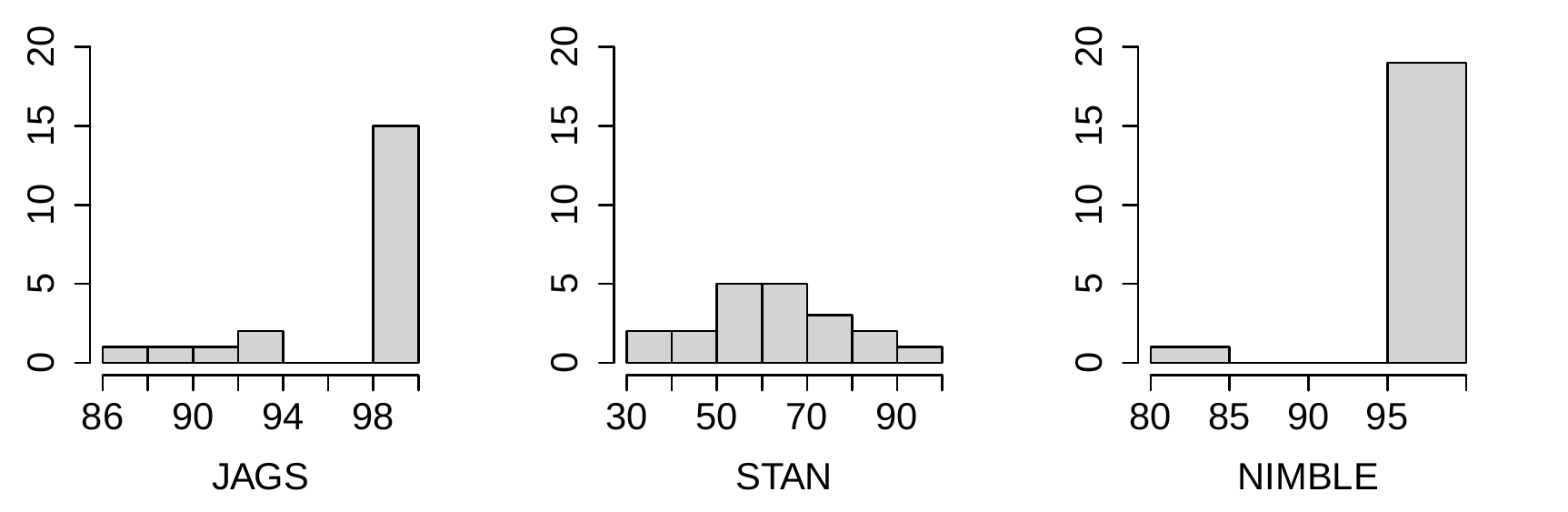}
			\caption{\textit{ess $v^2$}/$N_{s}$ expressed in \%.}
			
		\end{subfigure}
		\hfill
		\vspace{15pt}\\
		\begin{subfigure}{1\textwidth}
			\centering
			\includegraphics[width=0.66\textwidth]{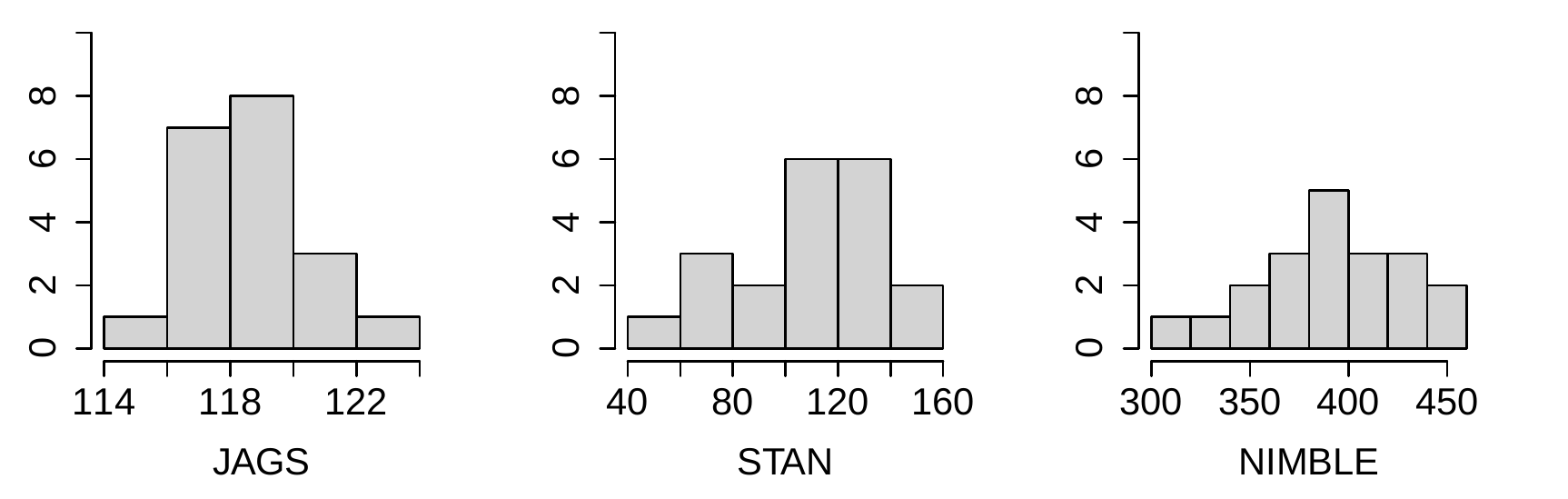}
			\caption{$N_{it}$/$t_{s}$}
			
		\end{subfigure}
		\caption{Histograms of \textit{ess$\beta$}/$N_{s}$ (a) and $N_{it}$/$t_{s}$ (b) over 20 simulated datasets ($n$=800, $H$=2) for the mixture models under prior \eqref{eq:mix_prior}.
}
		\label{fig:mixture_artiglieria}
	\end{figure}

\begin{figure}
		\centering
		\begin{subfigure}{1\textwidth}
			\centering
			\includegraphics[width=0.66\textwidth]{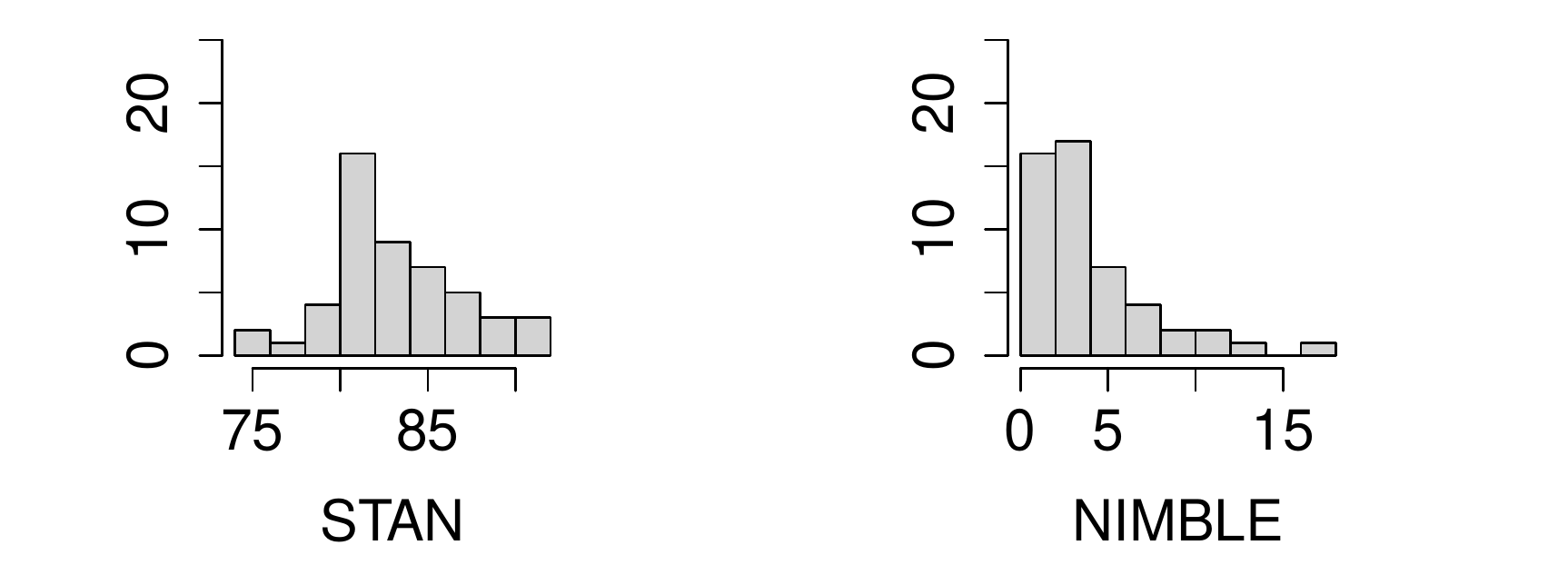}
			\caption{\textit{ess $\beta$}/$N_{s}$ expressed in \%.}
			
		\end{subfigure}
		\hfill
		\vspace{15pt}\\
		\begin{subfigure}{1\textwidth}
			\centering
			\includegraphics[width=0.66\textwidth]{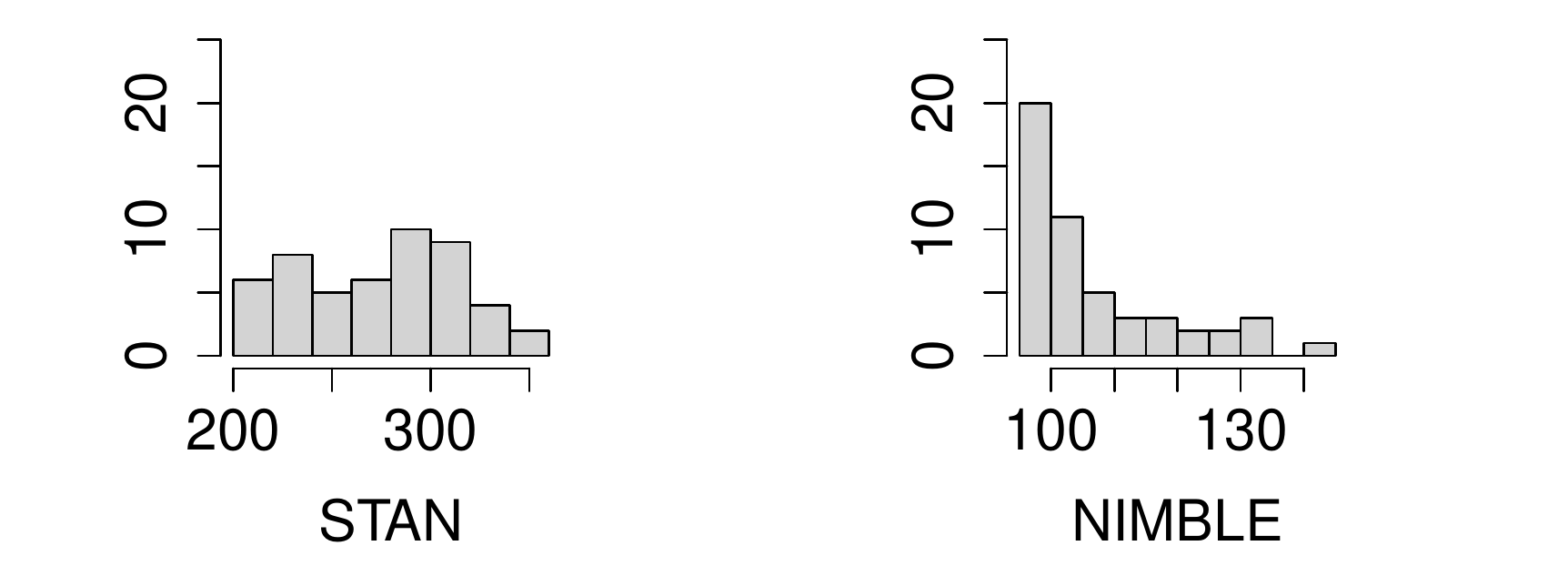}
			\caption{$N_{it}$/$t_{s}$}
			
		\end{subfigure}
		\caption{Histograms of \textit{ess$\beta$}/$N_{s}$ (a) and $N_{it}$/$t_{s}$ (b) over 50 simulated datasets ($n$=1000, $p$=8) for the AFT models under prior \eqref{eq:aft_nh}.
}
		\label{fig:aft_artiglieria}
	\end{figure}

Histograms of the monitoring indexes in Figures~\ref{fig:lm_conj_artiglieria}-\ref{fig:aft_artiglieria} confirm the conclusion from Section~\ref{sec:software_comparison}. For instance, Figure~\ref{fig:lm_conj_artiglieria} (a) displays the histogram of the values of the average of the \textit{ess} for all the $\beta_j$'s parameters, over the final sample size $N_s$, when we consider the different datasets.
For LMs under the conjugate prior (see Figure~\ref{fig:lm_conj_artiglieria}), JAGS and Stan always provide chains with high \textit{ess}, while NIMBLE generates highly autocorrelated chains.
Stan turns out to be slower than JAGS and NIMBLE in terms of the sampling time. Indeed, Stan generates around 350 samples per second, while JAGS and NIMBLE around 900 and 2500. 
On the other hand, if we consider LMs under the lasso prior, the logistic regression or AFTs, Stan is more efficient than JAGS and NIMBLE. Figure~\ref{fig:logreg_artiglieria} shows that in logistic regression models, the \textit{ess} of the chains generated by JAGS and NIMBLE is always smaller 30\%, while it is between 55\% and 90\% for Stan. In this case, Stan is faster than JAGS and NIMBLE, since it is able to generate around 400 samples per seconds, while JAGS and NIMBLE around 35 and 240 samples per second respectively. For LMs under the lasso prior and for AFT models (see Figures~\ref{fig:lm_lasso_artiglieria} and \ref{fig:aft_artiglieria}), JAGS has  not been considered in the comparison, due to very long sampling times. In both cases, the posterior MCMC chains generated by Stan are much less autocorrelated then those generated by NIMBLE.  Since, in addition,  
Stan is faster than NIMBLE, our final recommendation is for Stan
in the case of LM (under lasso prior), LR e AFT models.  
For MMs,  NIMBLE is much faster than JAGS and Stan, since it generates around 400 samples per second, while JAGS and Stan only around 120  (see Figure~\ref{fig:mixture_artiglieria}).
	
	Overall, it is clear that the findings we get from Section~\ref{sec:software_comparison} have been confirmed from this analysis with repeated simulated datasets. Summing up, our recommendation   is to rely on JAGS for conjugate or semi-conjugate linear models and on Stan for linear models under the lasso prior,  for logistic regression models and accelerated failure time models.  However, NIMBLE has proved to be  the most efficient software for mixture models.

	\FloatBarrier

\section*{Acknowledgements}
We are thankful to Giulia Gualtieri, Eugenia Villa and Riccardo Vitali, who contributed to an early version of most of the codes used in this manuscript.

	\section*{}
	\bibliography{references}
	\bibliographystyle{ba}
	
\end{document}